\documentclass[12pt]{article}
\usepackage{color}
\usepackage[dvips]{graphicx}
\usepackage{dcolumn}
\usepackage{amsmath}
\usepackage{url}
\begin{document}
% ----------------------------------------------------------
\Large
\begin{center}
	Practical Treatment of Singlet Oxygen with 
	Density-Functional Theory and the 
	Multiplet-Sum Method
\end{center}
\normalsize

\vspace{0.5cm}

\noindent
Abraham Ponra\\
{\em University of Maroua, Cameroon\\
e-mail: abraponra@yahoo.com}

\vspace{0.5cm}

\noindent
Anne Justine Etindele\\
{\em Higher Teachers Training College, University of Yaounde I,
P.O.\ Box 47, Yaounde, Cameroon\\
e-mail: annetindele@yahoo.fr}

\vspace{0.5cm}

\noindent
Ousmanou Motapon\\
{\em University of Maroua, Cameroon\\
e-mail: omotapon@univ-douala.com}

\vspace{0.5cm}

\noindent
Mark E.\ Casida\\
{\em Laboratoire de Spectrom\'etrie, Interactions et Chimie th\'eorique 
(SITh),
D\'epartement de Chimie Mol\'eculaire (DCM, UMR CNRS/UGA 5250),
Institut de Chimie Mol\'eculaire de Grenoble (ICMG, FR2607), 
Universit\'e Grenoble Alpes (UGA)
301 rue de la Chimie, BP 53, F-38041 Grenoble Cedex 9, FRANCE\\
e-mail: mark.casida@univ-grenoble-alpes.fr} 

% \vspace{0.5cm}
% 
% 
% \noindent
% {\color{magenta}   % begin magenta
% Date of Publication: \today \hspace{0.1cm} (MS 1.10)
% }                  % end magenta

\vspace{0.5cm}

\begin{center}
{\bf Abstract}
\end{center}
Singlet oxygen ($^1$O$_2$) comes in two flavors --- namely the dominant
lower-energy $a \,^1\Delta_g$ state and the higher-energy shorter-lived
$b \,^1\Sigma_g^+$ state --- and plays a key role in many photochemical
and photobiological reactions.  For this reason, and because of the 
large size of the systems treated, many papers have appeared with 
density-functional theory (DFT) treatments 
of the reactions of $^1$O$_2$ with different chemical species.  
The present work serves as a reminder that the common assumption
that it is enough to fix the spin multipicity as unity 
is not enough to insure a correct treatment 
of singlet oxygen.  We review the correct group theoretical treatment of the
three lowest energy electronic states of O$_2$ which, in the case
of $^1$O$_2$ is often so badly explained in the relevant photochemical
literature that the explanation borders on being incorrect and prevents, 
rather than encourages, a correct treatment of this interesting and 
important photochemical species. We then show how many electronic structure 
programs, such as a freely downloadable and personal-computer compatible
{\sc Linux} version of {\sc deMon2k}, may be used, together with the 
multiplet sum method (MSM), to obtain a more accurate estimation of the 
potential energy curves (PECs) of the two $^1$O$_2$ states. Applications of 
the MSM DFT method to $^1$O$_2$ appear to be extremely rare as we were only
able to find one correct application of the DFT MSM (or rather a very similar
approach) to $^1$O$_2$ in our literature search and then only using
a single functional.  Here we treat both the $a \,^1\Delta_g$ and 
$b \,^1\Sigma_g^+$ state with a wide variety of density-functional 
approximations (DFAs).  Various strengths and weaknesses 
of different DFAs emerge through our application of the MSM method.  In 
particular, the quality of the $a \,^1\Delta_g$ excitation energy reflects 
how well functionals are able to describe the spin-flip energy in DFT 
while the quality of the $b \,^1\Sigma_g^+$ excitation energy reflects how 
well functionals are able to describe the spin-pairing energy in DFT.  
Finally we note that improvements in DFT-based excited-state methods
will be needed to describe the full PECs of $^1$O$_2$ 
including both the equilibrium bond lengths and dissociation behavior.

% =================================================
\section{Introduction}
\label{sec:intro}

Modern chemistry is usually said to begin with the epoch
of pneumatic chemistry when chemists realized that, rather than a single
air, there are multiple airs (i.e., gases) which helped usher in the era
of modern chemistry.  Here ``modern chemistry'' may be roughly defined 
by Antoine-Laurent Lavoisier's famous book \cite{L1789} in which 
molecular oxygen, which makes up 20\% of our atmosphere, plays a 
particularly important role.  
Those of us who teach first-year 
University chemistry are well aware that explaining the paramagnetism of 
O$_2$ constitutes an early victory of molecular orbital (MO) theory.  
And yet, this diradical is not the highly-reactive species that experience 
with other radical species might lead us to believe.  Indeed there is an 
important kinetic barrier \cite{BHSC17} which prevents organic-based life 
forms such as we human beings from becoming interesting cases of spontaneous 
combustion.  Yet O$_2$ does react, with the right substances or given enough 
time or the right catalysts.  This is why many metals are only found in the 
Earth's crust as oxides.  Moreover O$_2$ is essential for respiration in 
animals which take in O$_2$ and give off CO$_2$ while plants produce O$_2$ 
from CO$_2$.  These examples indicate something about the richness of 
O$_2$ chemistry and imply both that much must already be known about this 
well-studied
molecule but also that there are good reasons why reactions of O$_2$ will 
continue to be studied for some time to come.  Molecular oxygen has three 
important low-lying states, namely the ground $X \,^3\Sigma_g^-$ triplet state 
($^3$O$_2$) and the two low-lying electronic states of singlet oxygen 
($^1$O$_2$), the lower $a \,^1\Delta_g$ state and the higher 
$b \,^1\Sigma_g^-$ state.  Several reviews have emphasized the importance 
of $^1$O$_2$ for photochemistry and photobiology 
\cite{K71,RT98,DC02,PCJO06,O10}.  Unfortunately a basic understanding of 
the fundamental quantum chemistry of the processes involved in $^1$O$_2$ 
reactions is hampered by the size of the systems interacting with 
$^1$O$_2$ which are typically treated with a computationally-simplified 
(even if theoretically-sophisticated) method such as density-functional 
theory (DFT).  Treating open-shell systems and excited-states with DFT 
is far from obvious for most users and so it is not surprising that 
almost all of the aforementioned applications are at best approximate 
(even conceptually incorrect).  This is also true for DFT treatments 
of isolated $^1$O$_2$.  It is the purpose of this paper to show how 
the Ziegler-Rauk-Baerends-Daul \cite{ZRB77,D94} 
multiplet sum method (MSM) provides a simple first approximation to a correct
treatment of $^1$O$_2$.  We investigated a number of functionals and found
which work best for this application.  We also discuss limitations of the 
MSM and alternatives based upon DFT which may be able to overcome
some of these limitations.

Much is already known about isolated $^1$O$_2$, making it a great molecule
for validating DFT methodology but no longer a molecule which is interesting
to study for its own sake.  However there is still
a great need to study how $^1$O$_2$ reacts with other molecules and how
$^1$O$_2$ is created by excited photosensitizers reacting with $^3$O$_2$.
Since we already have an interest in buckminsterfullerene (C$_{60}$) 
\cite{EMMM17,DCZ+21}, let us mention, as an example, that C$_{60}$ is of particular 
photochemical interest because it resists photodegradation and is not 
only a good electron acceptor but is also able to accept several electrons 
at a time.  Also C$_{60}$ is a photosensitizer \cite{DC02} for creating 
$^1$O$_2$ \cite{BLH+09,H18}.  Direct photoexcitation of $^3$O$_2$ 
to form $^1$O$_2$ is spectroscopically forbidden, but can be obtained by first
photoexciting a different molecule (i.e., the photosensitizer) which can
then transfer energy to $^3$O$_2$ to make $^1$O$_2$.  Yamakoshi {\em et al.}
showed experimentaly that C$_{60}$ can act as just such a photosensitizer
\cite{YUR+03} and this photosensitization process was later investigated
theoretically by Fueno {\em et al.} \cite{FTT11}.  Once formed $^1$O$_2$
is highly reactive and may react with many other molecules, including C$_{60}$.

% -----------------------------------------------------
\begin{table}
\caption{
\label{tab:compare}
Comparison of a few DFT $a \,^1\Delta_g$ and $b \,^1\Sigma_g^+$ excitation 
energies (Ha).  Differences between vertical and adiabatic excitation
energies (unspecified in this table) are small compared with the 
differences between the values obtained via different approaches.
}
\begin{center}
        \begin{tabular}{lccl}
\hline \hline
		Approach & $a \,^1\Delta_g$ & $b \,^1\Sigma_g^+$ & Reference \\
\hline
B3LYP/6-31G* &  0.0166$^a$ (0.0333$^b$) &                    & Ref.~\cite{GBOR98} \\
B3LYP/6-31G*  & 0.0166$^a$           &               & Ref.~\cite{FTT11} \\
B3LYP/6-311G+(d,p) &  0.0613$^a$ (0.033$^b$) &    & Ref.~\cite{AAG+17} \\
HSEH1PBE/6-311G++(3df,2p) & 0.0648$^a$ (0.032$^b$) &    & Ref.~\cite{AAG+17} \\
B3PW91/6-311G++(3df,2p)  & 0.0632$^a$ (0.032$^b$) &    & Ref.~\cite{AAG+17} \\
CAM-B3LYP/6-311G++(3df,2p) & 0.0621$^a$ (0.033$^b$) &    & Ref.~\cite{AAG+17} \\
M06/6-311G+(d,p) & 0.0594$^a$ (0.0503$^b$) &    & Ref.~\cite{AAG+17} \\
B97D3/TZVP S-T gap & 0.0642 &                    & Ref.~\cite{NISTCCCBDB} \\
BLYP/DEF2-TZVPP $\pi_x [\uparrow \downarrow] [\,\,\,\, \,\,\,\,] \pi_y$
		& 0.0568          &                    & PW$^c$ \\
BLYP/DEF2-TZVPP & 0.0273           & 0.0863            & PW$^c$ \\
M06/DEF2-TZVPP  & 0.0436           & 0.0674             & PW$^c$ \\
% M06HF/DEF2-TZVPP & 0.0292        &  0.06736           & PW$^c$ \\
MSDFT/M06-HF/cc-pVTZ  & 0.0430           & 0.0622             & Ref.~\cite{Q20} \\
Best Estimate & 0.0361           & 0.0658             & Ref.~\cite{FCY+14}   \\
\hline \hline
\end{tabular}
\end{center}
$^a$ Broken-symmetry calculation.\\
$^b$ After spin projection using, for example, the method of Yamaguchi {\em et al.}\\
$^c$ Present work.
\end{table}
% -----------------------------------------------------
The size of the C$_{60}$ + O$_2$ system combined with the accuracy and
cost efficiency that we have come to expect from DFT means that DFT is 
one of the most common methods used to study this and similar problems.
First, however, DFT must be properly validated for treating systems of 
this type which include both open-shell molecules and excited states.
Very interestingly, although it is relatively easy to find DFT studies of 
the chemistry of $^1$O$_2$ reacting with other molecules, it is remarkably 
difficult to find benchmark studies validating different DFT methods 
for $^1$O$_2$.  That is why the present work focuses uniquely on validating
DFT for $^1$O$_2$ before going on (in some other paper) to extending this
methodology to $^1$O$_2$ + C$_{60}$.  In order to see more clearly why this 
is needed, we have gathered values that we have found, as well as a few 
from the present work, into {\bf Table \ref{tab:compare}}.  
The very large range of excitation energies seen in this table indicates
that something is not quite right.  Furthermore, besides the present
work, we find only one other calculation of the $b \,^1\Sigma_g^+$ excitation
energy\cite{Q20}.
The usual approach used in the literature is to fix the spin multiplicity
at one and carryout spin-unrestricted symmetry-broken DFT calculations. 
For example, this is the appraoch taken in the aforementioned article of 
Fueno {\em et al.} \cite{FTT11} who quote a a B3LYP excitation energy 
of 0.453 eV compared with a CASSCF value of 0.890 eV and an experimental 
value of 0.977 eV \cite{K71} for the $^1\Delta_g$ state while we will 
show that the B3LYP MSM calculations give a $^1\Delta_g$ excitation 
energy of 0.914 eV.  In an earlier study \cite{GBOR98}, 
Garavelli {\em et al.} found that this
value can be very much improved by spin-projection using the Yamaguchi 
method \cite{YJDH88}.  However our own experience with symmetry breaking
in H$_2$ (unpublished) is that one should not count on 
the Yamaguchi spin-projection method to improve results unless the
spin-contamination is small as it can also make them worse.  We are not 
alone in finding difficulties with the Yamaguchi spin-projection 
method \cite{YKNY94,WS96}.  Nevertheless it is also the approach 
used in Ref.~\cite{AAG+17} (see the Supplementary Information for that
article).
The American National Institute of Standards and Technology (NIST)
Computational Chemistry Comparison and Benchmark Data Base (CCCBDB) 
gives a value of the O$_2$ singlet-triplet gap of 0.064204 Ha calculated 
at the B97D3/TZVP level \cite{NISTCCCBDB}.  This is clearly a very bad 
estimate of a true $^3$O$_2$ $\rightarrow$ $^1$O$_2$ excitation energy 
but it does compare well with our calculations of the 
$\pi_x [\uparrow \,\,\,\,][\uparrow \,\,\,\,]\pi_y$
$\rightarrow$ $\pi_x [\uparrow \downarrow] [\,\,\,\, \,\,\,\,] \pi_y$ 
energy (e.g., 0.05685 Ha with the BLYP functional.)
In fact, besides our own, we know of only one study \cite{Q20} which
gives $^1$O$_2$ energies of a similar quality to those of the MSM and
this is because the multistate DFT (MSDFT) method used there, though 
different from the MSM, has incorporated some aspects of the MSM.  
As the focus of that article was different from our own, only one 
functional was used.  Here we show that different choices of density 
functionals can lead to significant differences in the quality of 
DFT MSM calculated $^1$O$_2$ energies and potential energy curves (PECs), 
so that care in chosing the functional is important and compromises will 
likely have to be made.   Fortunately studies such as the present one can help
in making the choices and compromises.

We have yet another objective in writing this paper.  Every research paper
should be at least a little bit pedagogical in so far as the researcher
needs to teach the reader what he or she has learned.  In the present case,
we wish to go a little further by showing an example of quality research
based upon solid theoretical ideas but performed with inexpensive 
widely-available materials.  Thus all of the calculations
reported here could have been done (and should be reproducible) by anyone 
running {\sc Linux} using a freely downloadable version of the {\sc deMon2k} 
program \cite{GCC+12}.

This article is organized as follows: The electronic structure of $^1$O$_2$
is reviewed in the next section, including the dissociation problem, and
the MSM is presented.  Section~\ref{sec:details} presents key computational
details and results are presented and discussed in Sec.~\ref{sec:results}.
Section~\ref{sec:conclude} presents conclusions and a discussion of perspectives
for future work.  Many more
details are included in the {\em Supplementary Information} in order to make
our presentation particularly complete.

%%%%%%%
% EOF %
%%%%%%%
% -----------------------------------------------
\section{Theory}
\label{sec:theory}

% ----------------------------------------------------------------
\begin{figure}
\begin{center}
\includegraphics[width=0.8\textwidth]{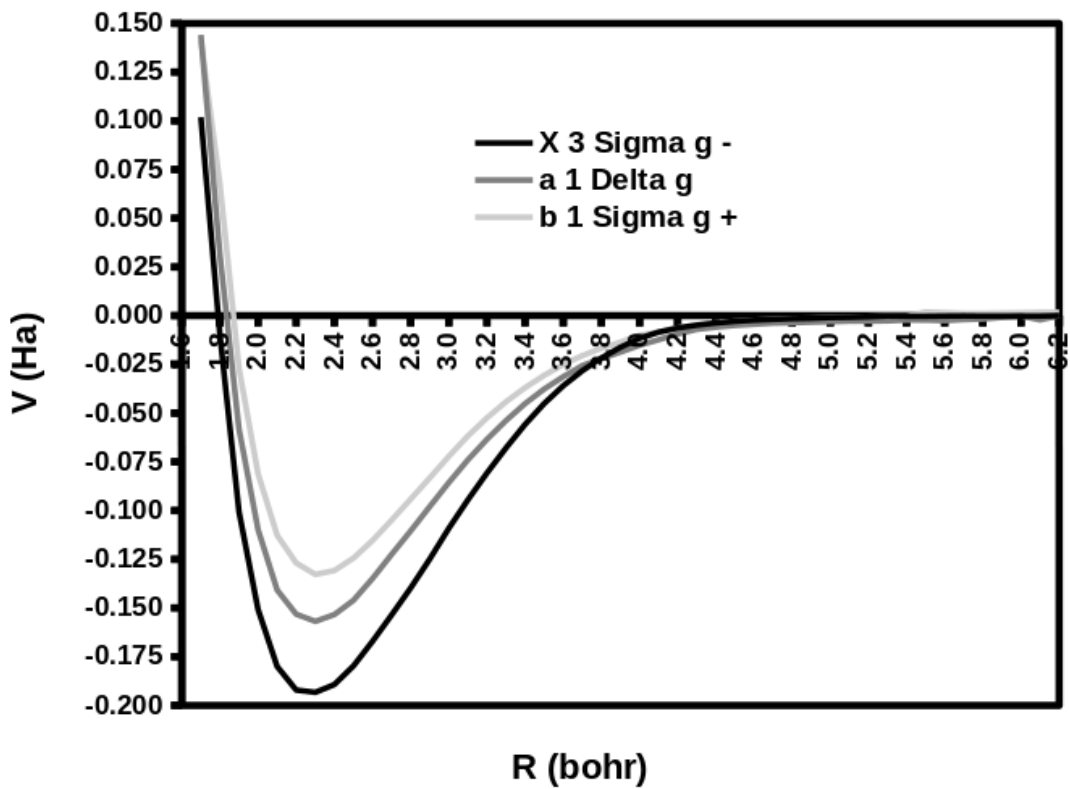}
\end{center}
\caption{
Potential energy curves for the three lowest electronic states
of O$_2$, namely $X \,^3\Sigma_g^-$, $a \,^1\Delta_g$, and $b \,^1\sigma_g^+$,
from Fig.~1 of Ref.~\cite{FCY+14} 
digitized with {\sc WebPlotDigitizer} \cite{wpd} and redrawn.  
The zero of energy is the twice the energy of the O($^3P$) atom.
\label{fig:O2PEC}
}
\end{figure}
% ----------------------------------------------------------------
We would like to use DFT to calculate the three lowest potential
energy curves (PECs) for O$_2$.  Accurate representations are
shown in {\bf Fig.~\ref{fig:O2PEC}}.  (See also Refs.~\cite{K72,SL77,VG02}.)
Specifically, these are the $^3\Sigma_g^-$ ground state, the 
lowest $^1\Delta_g$ singlet state, and a higher $^1\Sigma_g^+$ excited state.  
They arise from the different ways that two electrons may be placed into 
the two $\pi^*$ orbitals of the molecule.  In this section, we will first 
review the minimal configurations needed to describe these three states in
the context of wave function theory, then discuss the problem of correctly
dissociating O$_2$, and finally we will describe how the $^3\Sigma_g^-$,
$^1\Delta_g$, and $^1\Sigma_g^+$ may be calculated using DFT near the
equilibrium geometry.

% -----------------------------------------
\subsection{Wave function theory}
% -----------------------------------------

% -------------------------------------------------------------
\begin{figure}
\begin{center}
\includegraphics[width=0.8\textwidth]{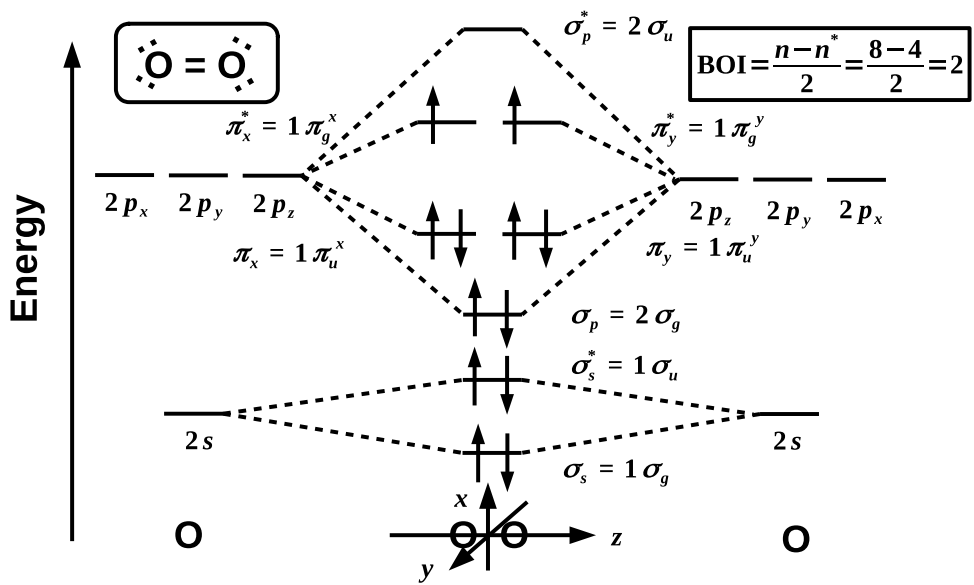}
\end{center}
\caption{\label{fig:O2}
First-year University description of the electronic structure of O$_2$:
atomic orbital/molecular orbital correlation diagram, Lewis dot structure 
(upper left inset), and calculation of the bond order index (BOI, upper 
right inset.)
}
\end{figure}
% --------------------------------------------------------------
{\bf Figure~\ref{fig:O2}} should be familiar.  It summarizes the molecular-orbital
(MO) picture of O$_2$ that most of us learn in our first-year University 
chemistry course.  It is a rather famous observation that, though the
Lewis dot structure seems to indicate that all electrons are paired, 
MO theory shows that the ground state of O$_2$ is, in fact, a triplet 
($^3$O$_2$).  Nevertheless, the calculation of the bond order index still 
shows a double bond.  

Low-lying excited states arise from the six symmetry-adapted linear 
combinations (SALCs) of the six different ways to put two electrons in
the two $\pi^*$ orbitals:
\begin{eqnarray}
  \pi_x^* [\uparrow \,\,\,\,] & \, & [\uparrow \,\,\,\,] \pi_y^*
  \nonumber \\ % 1
  \pi_x^* [\uparrow \downarrow] & \, & [\,\,\,\, \,\,\,\,] \pi_y^*
  \nonumber \\ % 2
  \pi_x^* [\uparrow \,\,\,\,] & \, & [\,\,\,\, \downarrow] \pi_y^*
  \nonumber \\ % 3
  \pi_x^* [\,\,\,\, \downarrow] & \, & [\uparrow \,\,\,\,] \pi_y^*
  \nonumber \\ % 4
  \pi_x^* [\,\,\,\, \,\,\,\,] & \, & [\uparrow \downarrow] \pi_y^*
  \nonumber \\ % 5
  \pi_x^* [\,\,\,\, \downarrow] & \, & [\,\,\,\, \downarrow] \pi_y^*
  \nonumber \\ % 6
  \label{eq:theory.1}
\end{eqnarray}
Making the SALCs requires the recognition that homonuclear diatomics like 
O$_2$ belong to the $D_{\infty h}$ point group whose 
character table is given in the {\em Supplementary Information}.
At the orbital level, the use of group
theory does not change very much, except that the MOs are now labeled as 
$g$ ({\em gerade}) or $u$ ({\em ungerade}) according to their inversion 
symmetry.  Figure~\ref{fig:O2} shows both the usual bonding and antibonding
and the group theoretic nomenclature for the MOs.  We will be using both
types of nomenclature.

% -------------------------------------------------------------
\begin{figure}
\begin{center}
\includegraphics[width=0.8\textwidth]{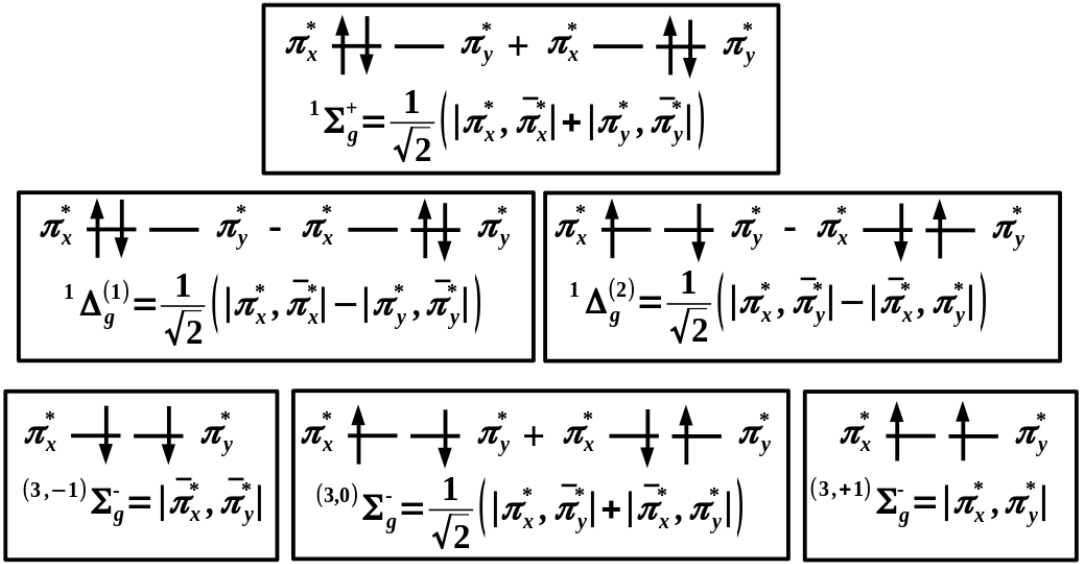}
\end{center}
\caption{\label{fig:MOpi_xy}
Practical representation of the lowest three electronic states of O$_2$
using real-valued MOs.
}
\end{figure}
% --------------------------------------------------------------
The many-electron state symmetry is determined by the orbitals outside
the closed-shell --- that is, by linear combinations of determinants
of with different occupations of the $\pi_x^* = \pi_g^x$  and 
$\pi_y^* = \pi_g^y$.  (See the {\em Supplementary Information} for a group 
theoretic construction of the SALCs.)  The result is summarized in 
{\bf Fig.~\ref{fig:MOpi_xy}} and is the representation that we will use
in the present work.

% -------------------------------------------------------------
\begin{figure}
\begin{center}
\includegraphics[width=0.8\textwidth]{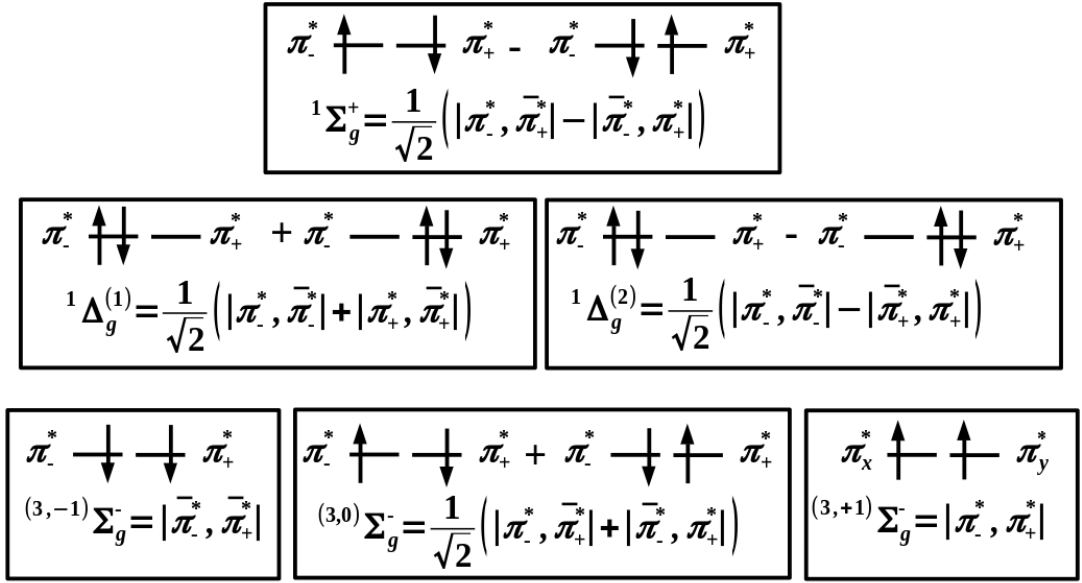}
\end{center}
\caption{\label{fig:MOpi_pm}
Conventional representation of the lowest three electronic states of O$_2$
using complex-valued MOs.
}
\end{figure}
% --------------------------------------------------------------
However this is not the representation which is used in most of the
literature.  Instead, the conventional analysis for O$_2$ 
(see, e.g., Ref.~\cite{K71}) uses the complex-valued orbitals, 
\begin{eqnarray}
  \pi_-^* & = & \frac{1}{\sqrt{2}} \left( \pi_x^* - i\pi_y^* \right) \nonumber \\
  \pi_+^* & = & \frac{1}{\sqrt{2}} \left( \pi_x^* + i\pi_y^* \right) \, .
  \label{eq:theory.2}
\end{eqnarray}
This leads to the conventional representation of the lowest three electronic
states shown in {\bf Fig.~\ref{fig:MOpi_pm}}.  Note that the MOs in 
Figs.~\ref{fig:MOpi_pm} and \ref{fig:MOpi_xy} differ by phase factors:
\begin{eqnarray}
  \vert \pi^*_- , \pi^*_+ \vert & = & i \vert \pi^*_x , \pi^*_y \vert \nonumber \\
	\frac{1}{\sqrt{2}} \left( \vert \pi^*_- , \bar{\pi}^*_+ \vert 
	+ \vert \bar{\pi}^*_- , \pi^*_+ \vert \right) & = & \frac{i}{\sqrt{2}} 
	\left( \vert \pi^*_x , \bar{\pi}^*_y \vert + 
	\vert \bar{\pi}^*_x , \pi^*_y \vert \right) 
	\nonumber \\
	\vert \bar{\pi}^*_- , \bar{\pi}^*_+ \vert & = & i \vert \bar{\pi}^*_x , \bar{\pi}^*_y \vert \nonumber \\
	\frac{1}{\sqrt{2}} \left( \vert \pi^*_- , \bar{\pi}^*_- \vert +
	\vert \pi^*_+ , \bar{\pi}^*_- \vert \right) & = & \frac{1}{\sqrt{2}} 
	\left( \vert \pi_x^* , \bar{\pi}_x^* \vert - \vert \pi^*_y , \bar{\pi}^*_y \vert \right)
	\nonumber \\
	\frac{1}{\sqrt{2}} \left( \vert \pi^*_- , \bar{\pi}^*_- \vert -
        \vert \pi^*_+ , \bar{\pi}^*_- \vert \right) & = &
	\frac{i}{\sqrt{2}} 
	\left( \vert \bar{\pi}^*_x , \pi^*_y \vert - \vert \pi^*_x , \bar{\pi}_y^* \vert \right)
	\nonumber \\
	\frac{1}{\sqrt{2}} \left( \vert \pi^*_- , \bar{\pi}^*_+ \vert - 
	\vert \bar{\pi}^*_- , \pi^*_+ \vert \right) & = & 
	\frac{1}{\sqrt{2}} \left( \vert \pi^*_x , \bar{\pi}_x^* \vert +
	\vert \pi^*_y , \bar{\pi}^*_y \vert \right) \, ,
  \label{eq:theory.3}
\end{eqnarray}
but are otherwise equivalent.  Unfortunately, though this convention is 
widely used \cite{L89,DC02,O10,WikiSingletOxygen}, it is now also 
only rarely explained and the $\pi$ orbitals are hardly ever properly labeled 
(see however Ref.~\cite{C06,PCJO06}).  We will use the real $\pi_x^*$ and $\pi_y^*$
MOs in the present work, rather than the complex valued $\pi_-^*$ and
$\pi_+^*$ MOs, as most quantum chemistry programs use real MOs.

% -----------------------------------------
\subsection{Dissociation limit}
\label{sec:theoryDL}
% -----------------------------------------

% -------------------------------------------------------------
\begin{table}
\caption{
\label{tab:terms}
Term symbols and relative energies of $2p^4$ multiplets of the oxygen atom
from the Atomic Spectra Database of the American National Institute of 
Standards and Technology \cite{NISTASD}.
}
\begin{center}
        \begin{tabular}{ccrr}
\hline \hline
        Term & $J$ & \multicolumn{2}{c}{Energy} \\
                &     & \multicolumn{1}{c}{cm$^{-1}$} & \multicolumn{1}{c}{eV} \\
\hline
$^1S$ & 0 & 33 792.483 & 4.1897359 \\
$^1D$ & 2 & 15 867.862 & 1.9673651 \\
$^3P$ & 0 & 226.977 & 0.0281416 \\
      & 1 & 158.265 & 0.0196224 \\
      & 2 & 0.000   & 0.0000000 \\
\hline \hline
\end{tabular}
\end{center}
\end{table}
% -------------------------------------------------------------
As we will be calculating PECs, we should also look at the dissociation
limit.  A prelude to this is to work out the possible
atomic term symbols.  There are many ways to do this, including, for example,
Hyde's very teachable double ladder method \cite{H75}.  The result may be
expressed in terms of Russell-Saunders coupling of the spin and orbital
magnetic moments. The term symbol takes the form $^{2S+1}L_J$ where 
\begin{equation}
  J = L+S, L+S-1, \ldots, \vert L-S \vert \, ,
  \label{eq:theory.4}
\end{equation}
is a new label used to distinguish the different levels with the same $L$ and
$S$.  Whichever way is used to derive the atomic term symbols, the result is
that shown in {\bf Table~\ref{tab:terms}}.  Hund's rules tell us that the ground 
state of the oxygen atom should be $^3P$.  Although Hund's rules do not
in general apply to excited states, they do give the proper ordering
of atomic states in the present case even for the excited states listed
in the table.

Let us now consider the problem of dissociating O$_2$.  According to 
Ref.~\cite{FCY+14} (and other sources) the $^3\Sigma_g^-$, $^1\Delta_g$,
and $^1\Sigma_g^+$ states all dissociate into two $^3P$ atoms.
Let us try to figure out how this works for the $^3\Sigma_g^-$ state.
This is basically an exercise in how molecular orbital theory is connected
with valence bond theory which can be represented by Lewis representations.
As we will be putting linear combinations of Slater determinants inside
Slater determinants, we should take a moment to clarify what is meant by 
a determinant within a determinant.  As the determinant of a matrix and the
determinant of the transpose of a determinant are the same, we could define
a Slater determinant either by,
\begin{equation}
  \vert \psi_1 , \psi_2 , \cdots , \psi_N \vert = \frac{1}{\sqrt{N!}}
  \sum_{\sigma \in {\cal S}_N} (-1)^\sigma \psi_{\sigma(1)}(1)
  \psi_{\sigma(2)}(2) \cdots \psi_{\sigma(N)}(N) \, ,
  \label{eq:theory.5}
\end{equation}
or by,
\begin{equation}
  \vert \psi_1 , \psi_2 , \cdots , \psi_N \vert = \frac{1}{\sqrt{N!}}
	\sum_{\sigma \in {\cal S}_N} (-1)^\sigma \psi_1(\sigma(1))
	\psi_2(\sigma(2) \cdots \psi_N(\sigma(N) \, ,
  \label{eq:theory.6}
\end{equation}
where ${\cal S}_N$ is the symmetric group for (i.e., the group of permutation 
of) the numbers $1,2,\cdots, N$ and $(-1)^\sigma$ is the sign of the 
permutation.  Only the second definition of the Slater determinant
is suitable for a general function.  Thus we may write that,
\begin{equation}
  \vert f(1,2,\cdots,N) \vert = \frac{1}{\sqrt{N!}}
        \sum_{\sigma \in {\cal S}_N} (-1)^\sigma
	f(\sigma(1), \sigma(2), \cdots , \sigma(N)) \, .
  \label{eq:theory.7}
\end{equation}
It is then a relatively simple exercise in group theory to show that,
\begin{equation}
  \vert \vert \psi_1, \psi_2, \cdots, \psi_P \vert , \psi_{P+1} ,\psi_N \vert
  = \sqrt{P!} \vert \psi_1, \psi_2, \cdots \psi_N \vert \, .
  \label{eq:theory.8}
\end{equation}
We also need that determinants are linear so that, for example,
\begin{equation}
  \vert \psi_1, \psi_2, \cdots, C_a \psi_a + C_b \psi_b , \cdots, \psi_N \vert
  = C_a \vert \psi_1, \psi_2, \cdots, \psi_a , \cdots \psi_N \vert 
  + C_a \vert \psi_1, \psi_2, \cdots, \psi_b , \cdots \psi_N \vert 
  \, .
  \label{eq:theory.9}
\end{equation}
This latter allows us to replace $p_{\pm 1}$ with $p_x$ and $p_y$ using the
relations,
\begin{eqnarray}
	p_x & = & \frac{1}{\sqrt{2}} \left( p_{+1}+p_{-1} \right) \nonumber \\
	p_y & = & \frac{1}{i\sqrt{2}} \left( p_{+1}-p_{-1} \right) \, .
  \label{eq:theory.10}
\end{eqnarray}
In particular,
\begin{equation}
  \vert p_x, p_y \vert = +i \vert p_{+1} , p_{-1} \vert
  \label{eq:theory.11}
\end{equation}
differ only by a phase factor.  Recall also that $p_0 = p_z$.

We may now make a connection with chemical bonding and in particular with
chemical bond breaking.  Orbitals $\chi_A$ and $\chi_B$ on two different
centers $A$ and $B$ may form bonding and antibonding combinations,
\begin{equation}
  \psi_{\pm} = \frac{1}{\sqrt{2}} \left( \chi_A \pm \chi_B \right) \, .
  \label{eq:theory.12}
\end{equation}
Here we have assumed that we are only interested in the dissociation limit
when the two centers are so far apart that any overlap between $\chi_A$
and $\chi_B$ is large enough to be able to define what is meant between
bonding and antibonding but small enough to be neglected in the normalization.
Note also that which of the plus and minus combinations is bonding and which 
is antibonding depends upon the precise nature of the orbitals $\chi_A$ and
$\chi_B$.  Filling both the bonding and antibonding combinations is equivalent
to bond breaking because,
\begin{eqnarray}
  \vert \psi_+ , \psi_- \vert = -\vert \chi_A , \chi_B \vert \, .
  \label{eq:theory.13}
\end{eqnarray}
This is why these terms cancel when calculating the bond order index 
(BOI = $(n-n^*)/2$, where $n$ is the number of electrons in bonding
orbitals and $n^*$ is the number of electrons in antibonding orbitals.)
Referring back to Fig.~\ref{fig:O2}, we see that we already have in the
dissociation limit that, dropping unimportant phase factors,
\begin{equation}
	\vert \sigma, \bar{\sigma}, \sigma^*, \bar{\sigma}^* , 
	\sigma_p , \bar{\sigma}_p , \pi_x, \bar{\pi}_x , \pi_y , \bar{\pi_y} \,
	\pi_x^* , \pi_y^* \vert \rightarrow 
	\vert s_A , \bar{s}_A, s_B, \bar{s}_B , \sigma_p , \bar{\sigma}_p ,
	p_x^A , p_x^B, p_y^A, p_y^B, \bar{\pi}_x , \bar{\pi}_y \vert
	\, .
  \label{eq:theory.14}
\end{equation}
Dissociating the $\sigma_p$ bond is a bit more complicated.  As the diatomic 
molecule traditionally lies along the $z$-axis, we have that
\begin{eqnarray}
  \sigma_p & = & \frac{1}{\sqrt{2}} \left( p_z^A - p_z^B \right) \nonumber \\
  \sigma_p^* & = & \frac{1}{\sqrt{2}} \left( p_z^A + p_z^B \right) \ .
  \label{eq:theory.15}
\end{eqnarray}
This leads to 
\begin{equation}
	\frac{1}{\sqrt{2}} \left( -\vert \sigma_p , \bar{\sigma}_p \vert
        + \vert \sigma^*_p , \bar{\sigma}^*_p \vert \right)
	= \frac{1}{\sqrt{2}} \left( \vert p_z^A , \bar{p}_z^B \vert
	- \vert \bar{p}_z^A , p_z^B \vert \right) \, ,
  \label{eq:theory.16}
\end{equation}
which is proper dissociation.  Hence (dropping phase factors)
\begin{eqnarray}
	& & 
	\frac{1}{\sqrt{2}} \vert \sigma, \bar{\sigma}, \sigma^*, \bar{\sigma}^* , 
	\frac{1}{\sqrt{2}} \left( -\vert \sigma_p , \bar{\sigma}_p \vert 
	+ \vert \sigma^*_p , \bar{\sigma}^*_p \vert \right) , \pi_x, 
	\bar{\pi}_x , \pi_y , \bar{\pi_y} ,
	\pi_x^* , \pi_y^* \vert \nonumber \\
	& \rightarrow  &   \frac{1}{\sqrt{2}}
	\vert s_A , \bar{s}_A, s_B, \bar{s}_B , p_z^A , \bar{p}_z^B ,
	p_x^A , p_x^B, p_y^A, p_y^B, \bar{\pi}_x , \bar{\pi}_y \vert
	\nonumber \\
	& - &  \frac{1}{\sqrt{2}} \vert s_A , \bar{s}_A, s_B, \bar{s}_B , 
	\bar{p}_z^A , p_z^B ,
	p_x^A , p_x^B, p_y^A, p_y^B, \bar{\pi}_x , \bar{\pi}_y \vert 
	\, .
  \label{eq:theory.17}
\end{eqnarray}
Getting proper atomic dissociation into $^3P$ components requires us to
use that,
\begin{eqnarray}
	\bar{\pi}_x & = & \bar{p}_x^A + \bar{p}_x^B \nonumber\\
	\bar{\pi}^*_x & = & \bar{p}_x^A - \bar{p}_x^B \nonumber \\
	\bar{\pi}_y & = & \bar{p}_y^A + \bar{p}_y^B \nonumber \\
	\bar{\pi}^*_y & = & \bar{p}_y^A - \bar{p}_y^B \, .
  \label{eq:theory.18}
\end{eqnarray}
Then
\begin{equation}
	\frac{1}{\sqrt{2}} \left( \vert \bar{\pi}_x, \bar{\pi}_y \vert
	- \vert \bar{\pi}_x^*, \bar{\pi}^*_y \vert \right) =
	\frac{1}{\sqrt{2}} \left( \vert \bar{p}_x^A , \bar{p}_y^B \vert
	+ \vert \bar{p}_y^A, \bar{p}_x^B \vert \right) \, .
  \label{eq:theory.19}
\end{equation}
Hence
\begin{eqnarray}
	& & 
	\frac{1}{2} \vert \sigma, \bar{\sigma}, \sigma^*, \bar{\sigma}^* , 
	\frac{1}{\sqrt{2}} \left( -\vert \sigma_p , \bar{\sigma}_p \vert 
	+ \vert \sigma^*_p , \bar{\sigma}^*_p \vert \right) , \pi_x, 
	\bar{\pi}_x , \pi_y , \bar{\pi_y} ,
	\frac{1}{\sqrt{2}} \left( \vert \bar{\pi}_x, \bar{\pi}_y \vert
	- \vert \bar{\pi}^*_x , \bar{\pi}^*_y \vert \right) \vert \nonumber \\
	& \rightarrow  &   \frac{1}{2}
	\vert s_A , \bar{s}_A, s_B, \bar{s}_B , p_z^A , \bar{p}_z^B ,
	p_x^A , p_x^B, p_y^A, p_y^B, \bar{p}_x^A , \bar{p}_y^B \vert
	\nonumber \\
	& + & \frac{1}{2}
        \vert s_A , \bar{s}_A, s_B, \bar{s}_B , p_z^A , \bar{p}_z^B ,
        p_x^A , p_x^B, p_y^A, p_y^B, \bar{p}_y^A , \bar{p}_x^B \vert
        \nonumber \\
	& - &  \frac{1}{2} \vert s_A , \bar{s}_A, s_B, \bar{s}_B , 
	\bar{p}_z^a , p_z^B ,
	p_x^A , p_x^B, p_y^A, p_y^B, \bar{p}_x^A , \bar{p}_y^B \vert 
        \nonumber \\
	& - &  \frac{1}{2} \vert s_A , \bar{s}_A, s_B, \bar{s}_B , 
	\bar{p}_z^A , p_z^B ,
	p_x^A , p_x^B, p_y^A, p_y^B, \bar{p}_y^A , \bar{p}_x^B \vert 
	\, ,
  \label{eq:theory.20}
\end{eqnarray}
which is a sum of four terms each corresponding to two $^3P$ oxygen atoms.  
(Compare with Fig.~43 of Ref.~\cite{SCZ+19}.)

The only wave function in ordinary DFT is that of the reference system of
non-interacting electrons which is single-determinantal for a non-degenerate
system.  This is why DFT is assumed to work best for systems which can
be described to a first approximation by a single-determinantal wave
function.  This is clearly not the case for the dissociation limit of 
the $^3$O$_2$ ground state.  However, as is commonly illustrated for
H$_2$, we may often obtain a reasonable approximation to the ground state
PEC by symmetry breaking in spin-unrestricted DFT calculations.  Hence,
for $^3$O$_2$ we can imagine calculating the PEC by symmetry breaking
and using a 
$\vert s_A , \bar{s}_A, s_B, \bar{s}_B , p_z^A , \bar{p}_z^B ,
p_x^A , p_x^B, p_y^A, p_y^B, \bar{p}_y^A , \bar{p}_x^B \vert$ (i.e.,
$\vert s_A , \bar{s}_A , p_x^A , p_y^A , \bar{p}_y^A , p_z^A \vert$ and
$\vert s_B, \bar{s}_B , p_x^B, \bar{p}_x^B , p_y^B, \bar{p}_z^B$.)

We should emphasize the difficulties with applying symmetry breaking
in this case.  As molecules become more complex, there may be more than
one way to lower the energy by symmetry breaking.  Finding the different
ways to break symmetry and choosing the best one is, as far as we know,
an unsolved problem.  The problem of multiple ways to break symmetry
is likely to be relevant for $^3$O$_2$ as it is not enough to
break the $\sigma_p$ or $\pi_x$ or $\pi_y$ symmetries but we 
also have to break them all simultaneously in the right way, which is 
not at all an obvious thing to do.  

The situation is more severe when trying to use DFT to treat the dissociation
limit of excited states.  This is because the MOs no longer belong to
well-defined irreducible representations of the point group, making it 
difficult to assign the symmetries of any excited states made from these
MOs.

% ------------------------------------------------------
\subsection{Singlet States from the Multiplet Sum Method}
% ------------------------------------------------------

Although DFT seems to only work for states which may be described
by a single-determinant reference, we may obtain the energies of other 
states using the Ziegler-Rauk-Baerends (Daul) multiplet sum method 
(MSM) \cite{ZRB77,D94}.  
This works for single-point (i.e., at a 
single geometry) calculations but we are only aware of one implementation 
of analytic gradients which would allow geometry optimizations with 
this method Friedrichs and Frank used such a method in the {\sc CPMD} 
program to do excited-state dynamics \cite{FF09}.

To see how to apply the MSM, we will consider the three $^3\Sigma_g^-$
wavefunctions,
\begin{eqnarray}
  \,^{(3,1)}\Sigma_g^- & = & \vert \pi_x^*, \pi_y^* \vert \nonumber \\
  \,^{(3,0)}\Sigma_g^- & = & \frac{1}{\sqrt{2}} \left(
  \vert \pi^*_x , \bar{\pi}_y^* \vert + \vert \bar{\pi}_x^*, \pi_y^* \vert 
	\right) \nonumber \\
  \,^{(3,-1)}\Sigma_g^- & = & \vert \bar{\pi}_x^*, \bar{\pi}_y^* \vert \, ,
  \label{eq:theory.21}
\end{eqnarray}
and the two $^1\Delta_g$ wavefunctions,
\begin{eqnarray}
  \,^1\Delta_g^{(1)} & = & \frac{1}{\sqrt{2}} \left(
  \vert \pi_x^*, \bar{\pi}_x^* \vert - \vert \bar{\pi}_y^* , \pi_y^* \vert 
  \right) \nonumber \\
  \,^1\Delta_g^{(2)} & = & \frac{1}{\sqrt{2}} \left(
  \vert \pi_x^*, \bar{\pi}_y^* \vert - \vert \bar{\pi}_x^* , \pi_y^* \vert 
  \right) \, .
  \label{eq:theory.22}
\end{eqnarray}
Then
\begin{eqnarray}
	E[\,^3\Sigma_g^-] & = & E[\,^{(3,0)}\Sigma_g^-] 
	= E[\vert \pi_x^*,\bar{\pi}_y^* \vert] 
	+ \langle \vert \pi_x^*,\bar{\pi}_y^* \vert \, \vert \hat{H} \vert \,
	\vert \bar{\pi}_x^*, \pi_y^* \vert \rangle \nonumber \\
	E[\,^1\Delta_g] & = & E[\,^1\Delta_g^{(2)}] 
	= E[\vert \pi_x^*,\bar{\pi}_y^* \vert] 
	- \langle \vert \pi_x^*,\bar{\pi}_y^* \vert \, \vert \hat{H} \vert \,
	\vert \bar{\pi}_x^*, \pi_y^* \vert \rangle \, .
  \label{eq:theory.23}
\end{eqnarray}
Hence,
\begin{equation}
  E[\,^3\Sigma_g^-] + E[\,^1\Delta_g] = 2 E[\vert \pi_x^*,\bar{\pi}_y^* \vert]
  \, ,
  \label{eq:theory.24}
\end{equation}
and,
\begin{eqnarray}
   E[\,^1\Delta_g] & = & 2 E[\vert \pi_x^*,\bar{\pi}_y^* \vert]
   - E[\,^3\Sigma_g^-] \nonumber \\
   & = & 2 E[\vert \pi_x^*,\bar{\pi}_y^* \vert] 
	- E[\,^{(3,0)}\Sigma_g^-] \nonumber \\
	& = & 2 E[\vert \pi_x^*,\bar{\pi}_y^* \vert] 
	- E[\vert \pi_x^* , \pi_y^* \vert ] \, .
  \label{eq:theory.25}
\end{eqnarray}
Thus the $^1\Delta_g$ energy has been expressed purely in terms of the
energies of two single-determinantal states---namely the fictitious
mixed-symmetry $\vert \pi_x^*,\bar{\pi}_y^* \vert$ state and the
triplet state $\vert \pi_x^* , \pi_y^* \vert$.  
In principle, these should all be calculated using the same MOs
derived from some reference calculation (see below.)

In principle the extension of the MSM to the 
\begin{equation}
	\,^1\Sigma_g^+ = \frac{1}{\sqrt{2}} \left( \vert \pi_x^* , \bar{\pi}_x^* \vert + \vert \pi_y^* , \bar{\pi}_y^* \vert \right) 
  \label{eq:theory.26}
\end{equation}
is just as straightforward as
\begin{eqnarray}
  E[\,^1\Sigma_g^+] & = & E[\vert \pi_x^* , \bar{\pi}_x^* \vert]
  + \langle \vert \pi_x^* , \bar{\pi}_x^* \vert \, \vert \hat{H} \vert \,
  \vert \pi_y^* , \bar{\pi}_y^* \vert \rangle \nonumber \\
	E[\,^1\Delta_g] & = & E[\,^1\Delta_g^{(1)}]=
  E[\vert \pi_x^* , \bar{\pi}_x^* \vert]
  - \langle \vert \pi_x^* , \bar{\pi}_x^* \vert \, \vert \hat{H} \vert \,
  \vert \pi_y^* , \bar{\pi}_y^* \vert \rangle \, .
  \label{eq:theory.27}
\end{eqnarray}
So,
\begin{equation}
  E[\,^1\Sigma_g^+] + E[\,^1\Delta_g] = 2 E[\vert \pi_x^* , \bar{\pi}_x^* \vert]
  \, ,
  \label{eq:theory.28}
\end{equation}
and
\begin{equation}
  E[\,^1\Sigma_g^+] = 2 E[\vert \pi_x^* , \bar{\pi}_x^* \vert] - E[\,^1\Delta_g]
  \, .
  \label{eq:theory.29}
\end{equation}

For the reader's
convenience, we will repeat the key formulae again but this time in
a somewhat different and more physically more picturesque manner:
\begin{eqnarray}
  E[X \,^3\Sigma_g^-] & = &
  E(\pi_x^*[\uparrow, \,\,\,\,] [\uparrow , \,\,\,\,] \pi_y^*) \nonumber \\
  E[a \,^1\Delta_g] 
	& = & E(\pi_x^*[\uparrow, \,\,\,\,] [\uparrow, \,\,\,\,] \pi_y^*)
	+ 2 F \nonumber \\
  E[b \,^1\Sigma_g^+] & = &
  E(\pi_x^*[\uparrow, \,\,\,\,] [\uparrow, \,\,\,\,] \pi_y^*) + 2 P \, ,
  \label{eq:theory.30}
\end{eqnarray}
where the spin-flip energy,
\begin{equation}
  F = E(\pi_x^*[\uparrow, \,\,\,\,] [\,\,\,\, , \downarrow] \pi_y^*)
  - E(\pi_x^*[\uparrow, \,\,\,\,] [\uparrow, \,\,\,\,] \pi_y^*)
  \label{eq:theory.31}
\end{equation}
and the spin-pairing energy
\begin{equation}
  P = E(\pi_x^*[\uparrow, \downarrow] [\,\,\,\, , \,\,\,\,] \pi_y^*)
  - E(\pi_x^*[\uparrow, \,\,\,\,] [\,\,\,\, , \downarrow] \pi_y^*) \, .
  \label{eq:theory.32}
\end{equation}
This way of writing the equations tells us that the
$X \,^3\Sigma_g^- \rightarrow a \,^1\Delta_g$ excitation is going to
be sensitive to how well a given density-functional approximation (DFA)
can describe spin-flip energies and that the
$X \,^3\Sigma_g^- \rightarrow b \,^1\Sigma_g^+$
excitation is going to be sensitive to how well a given DFA can
describe spin-pairing energies.

% ----------------------------------
\subsection{Reference State Problem}
% ----------------------------------

As noted above, ideally the MSM should make use of a single set of MOs
coming from some reference calculation.  The choice of reference calculation
may be governed by several factors, including ease of convergence.  However,
it should be emphasized that the reason for using a reference state is
not just that some calculations converge more easily than others.  The
reference calculation also guarantees that the different determinants which 
are created are properly orthogonal to each other so as to avoid variational
collapse.  The same problem is addressed in complete active state 
self-consistent field (CAS-SCF) calculations by using state averaging.
Of course, the price that has to be paid is that orbital relaxation
is neglected, which is not always a good thing.  

Let us now turn our attention specifically to $^1$O$_2$.  The result 
of specifying a spin multiplicity of one and running a calculation 
is that most programs will try to pair up electrons by putting two 
electrons in a single $\pi^*$ orbital.  Such a doubly-occupied 
$\pi^*$ orbital will no longer be energetically degenerate with the 
other (empty) $\pi^*$ orbital.  In fact, the occupied $\pi^*$ orbital 
will have a higher energy than the unoccupied $\pi^*$ orbital
because of self-interaction errors.  Formal DFT predicts that this can happen
in open-shell systems even for the exact functional when noninteracting 
$v$-representability (NVR) fails.  We call this an effective failure of NVR
when it occurs for an approximate functional \cite{CH12}.  In either case,
the result is that most programs will try to satisfy the {\em Aufbau} principle
at each iteration by transfering electrons from the higher-energy occupied
$\pi^*$ orbital to the lower-energy unoccupied $\pi^*$ orbital.  As this
then raises the energy of the newly occupied orbital and lowers the energy
of the newly unoccupied orbital, we have an unstable situation and the 
calculations will typically no longer converge.

This problem may be solved by creating a reference state with each $\pi^*$
orbital having half a spin-up and half spin-down electron.  Most programs
have an option which allows the user to create this fractionally-occupied
state easily.  Here we used the {\sc deMon2k} program \cite{deMon2k,GCC+12}.  
In this program, it suffices to use the keyword combination
{\tt SMEAR 0.1 UNIFORM} where the number 0.1 is an adjustable number in eV
which controls how close orbitals are in energy before they are uniformly
occupied.  {\em A priori} this type of fractionally-occupied state corresponds
to some sort of ensemble electronic state, where the ensemble average has
been carried out over several multiplet states.  However deeper thought
indicates that this conclusion is far from obvious.
Another possibility (not tried) would be to use a carefully-selected 
restricted open-shell Kohn-Sham (ROKS) calculation as a reference state.
Once such a reference state is created, electrons may be displaced to create
different occupation states without additional iterations.  In {\sc deMon2k},
this is done with the keywords {\tt SCFTYPE UKS MAX=0} and {\tt MOMODIFY}.

%%%%%%%
% EOF %
%%%%%%%
% -----------------------------------------------
\section{Computational Details}
\label{sec:details}

This is a very short section as much of the computational method
has either been described in the Theory Section (Sec.~\ref{sec:theory}), 
in the {\em Supplementary Material}, or will be described in 
the Results Section (Sec.~\ref{sec:results}). We used 
a Gaussian-type orbital (GTO) based quantum chemistry code which
makes use of both GTO-type basis functions for expanding the 
molecular orbitals but also uses GTO-type basis functions to
expand the density (for local DFAs) and the density matrix (for 
hybrid DFAs) in order to eliminate four-center electron repulsion
integrals.  In particular, we used version 5.0 of {\sc deMon2k} 
which is freely downloadable from the web site and runs under 
{\sc Linux} \cite{deMon2k}.  The code is described in more detail
in Refs.~\cite{GCC+12,DAH+19}.  Our calculations used the DEF2-TZVPP orbital
basis set and the GEN-A3* auxiliary basis set.  Nearly all the functionals
available in this particular version of {\sc deMon2k} were used.  (No
Hartree calculations were performed.)
Literature citations for the specific functionals may be found in the Tables
in the {\em Supplementary Material}.  For other parameters,
we simply made extensive use of default options.  Occasionally we 
made use of convergence options other than the default when this seemed useful.
Atomic O($^3P$) energies were calculated at the same level as for
diatomic O$_2$.  No correction was made for basis set superposition
error.  Nearly all calculations were done without symmetry breaking and
a sample input file has been included in the {\em Supplementary Material}
for those who wish to reproduce our calculations.  Calculations with
symmetry breaking used a somewhat different procedure which is described 
in Sec.~\ref{sec:results} though our purpose in describing symmetry broken
calculations is primarily to caution against them.

%%%%%%%
% EOF %
%%%%%%%
% -----------------------------------------------
\section{Results}
\label{sec:results}

We wish to investigate the strengths and limitations of different
density-functional approximations (DFAs) for use with the multiplet sum
method (MSM) for treating the first three states ($X \,^1\Sigma_g^-$,
$a \,^1\Delta_g$, and $b \,^1\Sigma_g^+$) of O$_2$.  These calculations 
could be carried out with just about any DFT program, but we wish to 
be simultaneously both pedagogic and to show that quality research can
be done even with limited computer resources, so we have chosen to use
a freely downloadable and personal-computer compatible {\sc Linux} version 
of {\sc deMon2k}.  Also for pedagogical clarity, after briefly reviewing
our source of {\em best estimate} comparison data, 
we will spend a bit of time showing what happens when 
the method is applied in combination with symmetry breaking.  
This is important because many people try to deal with problems in
DFT through symmetry breaking and we wish to discuss why this fails
through a concrete example.  We will then go on to compare MSM calculations for 
O$_2$($X \,^1\Sigma_g^-$), O$_2$($a \,^1\Delta_g$), and 
O$_2$($b \,^1\Sigma_g^+$) without symmetry breaking using various functionals.
Finally we will choose one of the better functionals and compare our MSM
PECs with our reference PECs.
Unless otherwise specified, results will be given in Hartree atomic
units ($\hbar = m_e = e$).  This is simply convenient when interconverting
large amounts of data.  Results are summarized graphically here, but tables 
of calculated values may be found in the {\em Supplementary Information} where
references may also be found for the functionals used in the present paper.

% --------------------------------------------------
\subsection{Choice of Best Estimate Comparison Data}
% --------------------------------------------------

Because of the great importance of O$_2$, there are many reference studies ---
in fact, too many to discuss specifically here.  Some particularly notable
reviews are Refs.~\cite{K72,SL77,VG02}.  We found it particularly convenient
to digitize data from Fig.~1 of Ref.~\cite{FCY+14}.  These data are 
given in the {\em Supplementary Information} and regraphed in 
Fig.~\ref{fig:O2PEC}.  Another useful source of comparison data is 
the American National Institute of Standards and Technologies (NIST) Atomic 
Spectra Database \cite{NISTASD} to which we will also sometimes refer.

% --------------------------------------------------
\subsection{With Symmetry Breaking}
% --------------------------------------------------

This works best in the case of the O$_2$($X \,^3\Sigma_g^-$) ground 
state and becomes almost nonsensical for the O$_2$($a \,^1\Delta$)
and O$_2$($b \,^1\Sigma_g^+$) excited states.

% ----------------------------------------------------------------
\begin{figure}
\begin{center}
\includegraphics[width=0.8\textwidth]{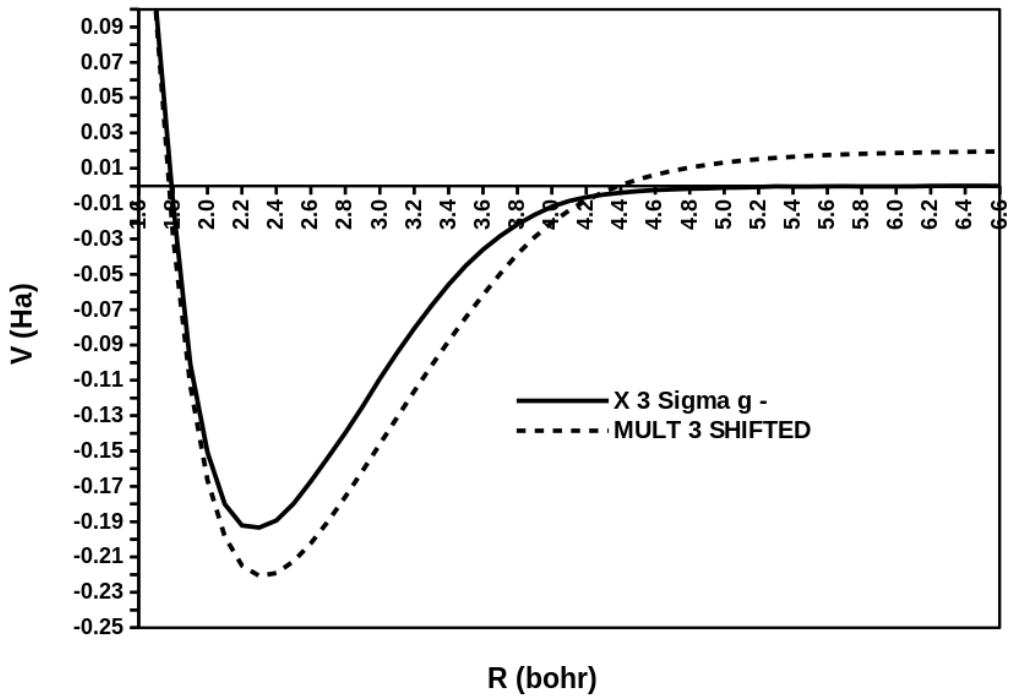}
\end{center}
\caption{
O$_2$($X \,^3\Sigma_g^-$) PECs calculated using the symmetry
breaking procedure described in the text: best estimate, solid line;
broken symmetry BLYP, dashed line.  The zero of energy is
that of twice the energy of a single O($^3P$) atom without
any basis set superposition error (BSSE) correction.
\label{fig:lesson6ans5}
}
\end{figure}
% ----------------------------------------------------------------
Let us begin with the ground O$_2$($X \,^3\Sigma_g^-$) state.
As discussed in Sec.~\ref{sec:theoryDL}, breaking symmetry to obtain the
correct dissociation limit of O$_2$ into two ground-state O($^3P$) atoms
is anything but obvious.  As it happened, it was possible to first generate
a restart file from a calculation for a single O($^3P$) atom and then 
use this as an initial guess for an SCF calculation at large internuclear
distance.  We then gradually reduced the O$_2$ bond length using the
the restart file from the previous (slightly longer) bond length as the
initial guess for our SCF calculation.  The results are shown in 
{\bf Fig.~\ref{fig:lesson6ans5}}.  The general shape of the BLYP PEC is 
correct but it dissociates to a higher energy than twice the energy of 
two oxygen atoms, suggesting that further symmetry breaking might be 
possible.  (Or perhaps this is simply the best that can be done without 
resorting to a multideterminantal approach?)  It is also clear that BLYP 
overbinds the molecule.

% ----------------------------------------------------------------
\begin{figure}
\begin{center}
\includegraphics[width=0.6\textwidth]{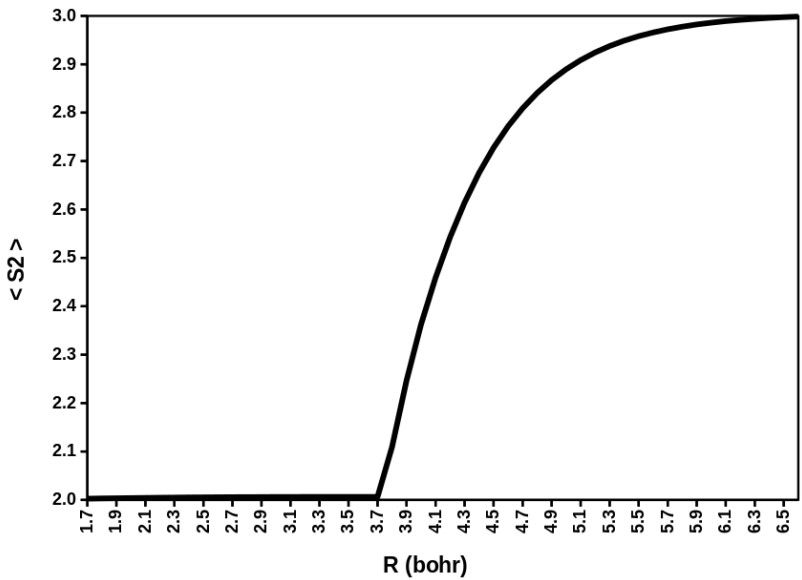}
\end{center}
\caption{
O$_2$($X \,^3\Sigma_g^-$) $\langle \hat{S}^2 \rangle$ for the symmetry
broken calculation.
\label{fig:lesson6ans6}
}
\end{figure}
% ----------------------------------------------------------------
Not all of the curve shown in Fig.~\ref{fig:lesson6ans5} is symmetry broken.
Symmetry breaking only occurs when there is a symmetry-broken 
[different-orbitals-for-different-spin (DODS)] solution which
is lower in energy than the symmetry-unbroken [same-orbitals-for-same-spin
(SODS)] solution.  One indication of when this occurs is where the value
of the expectation value of the spin operator $\langle \hat{S}^2 \rangle$
differs from the expected value for a triplet of $S(S+1)=2$.  As shown
in {\bf Fig.~\ref{fig:lesson6ans6}}, this occurs at a bond distance 
$R = \mbox{3.7 bohr}$. For larger values of $R$, $\langle \hat{S}^2 \rangle$
approaches the value of 3 which corresponds to a value of 
$S = (\sqrt{13}-1)/2 = 1.303$ which is not particularly physical.
When $R < \mbox{3.7 bohr}$, no symmetry breaking occurs (i.e., we have a
SODS calculation) and the PEC is the same as if no symmetry breaking at all
had occurred.  The exact value (also known as the Coulson-Fischer point) at
which symmetry breaking occurs will depend upon the choice of 
density-functional.  However we can expect that symmetry breaking will 
occur somewhere around $R = \mbox{3.7 bohr}$ for most density-functional
approximations (DFAs).

Before proceeding to treating the two excited $^1$O$_2$ states, we need
a reference state.  We have noticed that demanding a multiplicity of one
in many DFT programs results in the programming attempting to place the
two paired electrons in the {\em same} $\pi^*$ orbital which then has a
higher energy than the vacant $\pi^*$ orbital (because the filled $\pi^*$
orbital has an erroneous self-interaction energy not present in the vacant
$\pi^*$ orbital).  This results in the two paired electrons being dumped
into the other, lower energy $\pi^*$ orbital which then becomes higher
in energy and so forth, with the result that calculations do not converge.

% ----------------------------------------------------------------
\begin{figure}
\begin{center}
\includegraphics[width=0.8\textwidth]{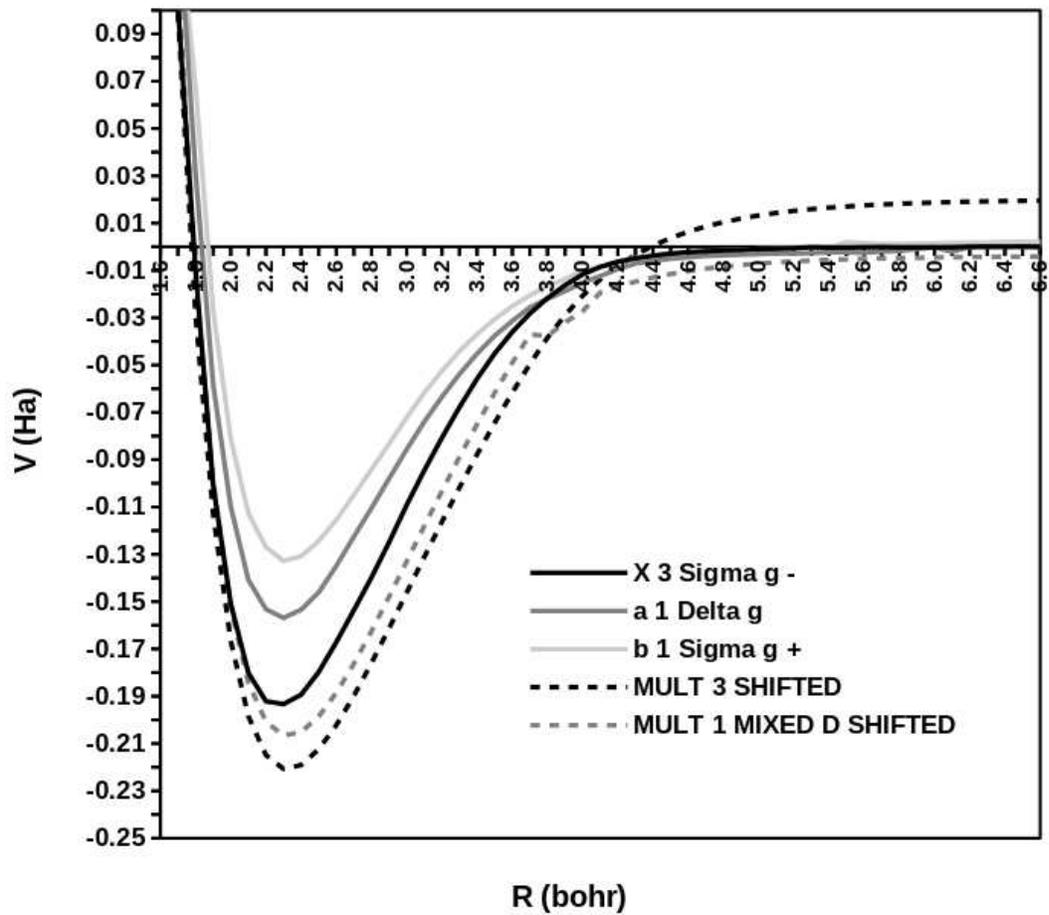}
\end{center}
\caption{
PEC for the $\pi_x^*[\,\,\,\, \,\,\,\,] [\uparrow \downarrow] \pi_y*$
reference state: solid lines, best estimates; black dashed line, broken symmetry
triplet; grey dashed line, broken symmetry reference.  
Note that the labels $\pi_x^*$ and $\pi_y^*$ are only
strictly correct when no symmetry breaking occurs.
\label{fig:lesson6ans7}
}
\end{figure}
\begin{figure}
\begin{center}
\includegraphics[width=0.8\textwidth]{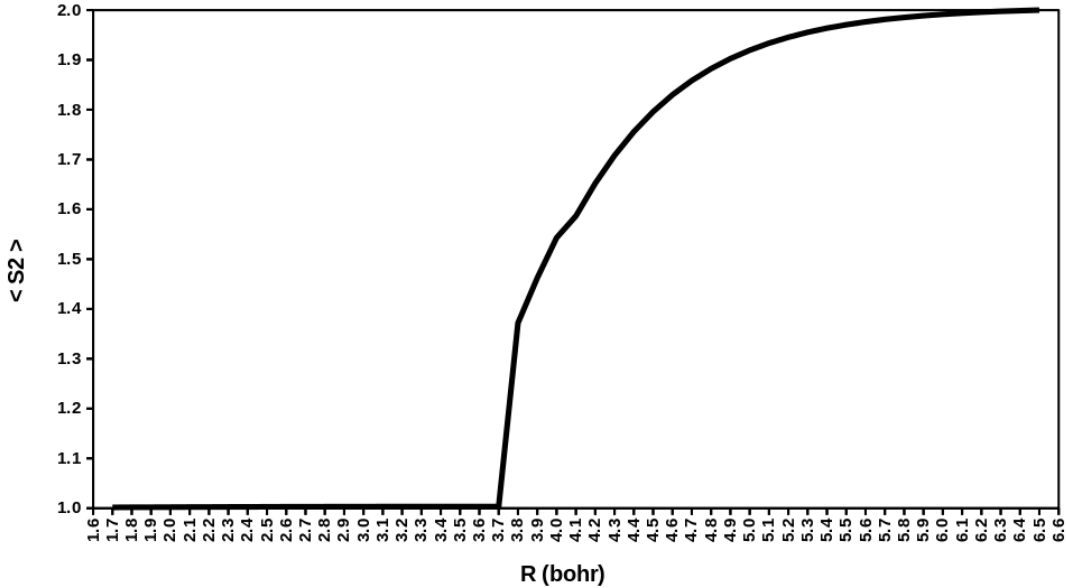}
\end{center}
\caption{
O$_2$ $\langle \hat{S}^2 \rangle$ for the reference state in the
symmetry-broken calculation.
\label{fig:lesson6ans8}
}
\end{figure}

% ----------------------------------------------------------------
However we found easy convergence to having the two paired electrons in
the same orbital in this case when we simply used the restart file from
our triplet calculation and imposed a spin multiplicity of one (keyword
{\tt MULTI 1}).  The PEC for this reference state and the BLYP DFA is 
shown in {\bf Fig.~\ref{fig:lesson6ans7}}.  
Notice the uneven behavior of the reference state PEC between 
$R = \mbox{3.7 \AA}$ (the Coulson-Fischer point) and $R = \mbox{4.1 \AA}$.
This might be a reflection of convergence difficulties which are often
encountered in the vicinity of Coulson-Fischer points.
{\bf Figure~\ref{fig:lesson6ans8}} shows
the spin-contamination in our spin-broken reference state as a function
of bond distance.  The value of $\langle \hat{S}^2 \rangle = S (S+1)$ 
should be zero for a singlet and two for a triplet.  The value of 
$\langle \hat{S}^2 \rangle$ is constant at $S(S+1)=1$ (corresponding
to an unphysical value of $S = (\sqrt{5}-1)/2 = 0.6180$) up to the same bond 
length where spin contamination set in for the triplet state.  It then 
seems to converge to a triplet state at very large values of $R$.
While this might seem strange for a calculation which fixes one up spin
and one down spin, it is perfectly consistent with the L\"owdin formula
\cite{L55} (Ref.~\cite{MC17} provides an ``easy'' derivation using
second quantization),
\begin{equation}
  \langle \hat{S}^2 \rangle = -\sum_{i,j} \vert \Delta_{i,j} \vert^2
  + \frac{n_\alpha + n_\beta}{2} + \left( \frac{n_\alpha - n_\beta}{2} \right)^2
  \, ,
  \label{eq:results.1}
\end{equation}
because $n_\alpha = n_\beta = 1$ and the spin-transfer matrix (${\bf \Delta}$) 
goes to zero at large distance in this case.
% ----------------------------------------------------------------
% ----------------------------------------------------------------

% ----------------------------------------------------------------
\begin{figure}
\begin{center}
\includegraphics[width=0.8\textwidth]{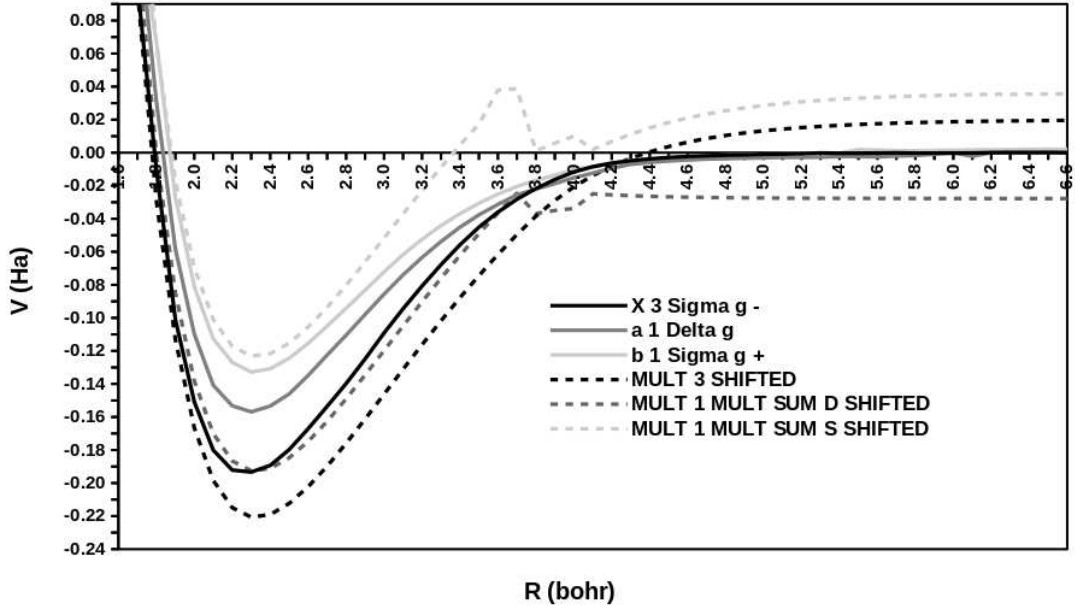}
\end{center}
\caption{
Comparison of O$_2$ MSM BLYP PECs from our symmetry-broken calculation
(dashed lines) with best estimate PECs (corresponding solid lines).
\label{fig:lesson6ans9}
}
\end{figure}
% ----------------------------------------------------------------
We may now use this reference state to carryout a MSM BLYP DFA calculation
whose results are shown in {\bf Fig.~\ref{fig:lesson6ans9}}.  Before
the Coulson-Fischer point at $R = \mbox{3.7 \AA}$, there is no symmetry
breaking and our calculation behaves much the same way as does the 
symmetry-unbroken calculations described below.  Once symmetry-breaking
sets in, the calculation is no longer straightforward.  In particular, 
one of the problems that we encountered in doing broken-symmetry MSM 
calculations is that the spin $\alpha = \, \uparrow$ MOs do not always 
correspond exactly with the spin $\beta = \, \downarrow$ MOs.  Usually this 
is just a matter of a reversal of orbital numbering.  However it could be 
more complicated and this inexact correspondance (together with the aforementioned
difficulties with the ground $X \,^3\Sigma_g^-$ state calculation) could explain the uneven 
behavior of the PECs between $R = \mbox{3.7 \AA}$ (the Coulson-Fischer point) 
and $R = \mbox{4.1 \AA}$.  Nevertheless we do obtain the correct ordering
of {\em singlet states} at large distance even though our results are certainly {\em not}
quantitatively correct and the triplet state lies incorrectly between the
two singlet states.  Of course, the most important difficulty and 
{\em the most important reason to avoid symmetry breaking when treating 
excited states} is that it is nearly impossible to construct excited states 
with well-defined symmetries unless the molecular orbitals belong to 
well-defined irreducible representations of the molecular point group, 
which is certainly not the case in broken-symmetry (DODS) calculations.
Henceforth we consider only symmetry-unbroken (SODS) calculations.

% --------------------------------------------------
\subsection{No Symmetry Breaking}
% --------------------------------------------------

We have calculated the PECs for the $X \,^3\Sigma_g^-$ ground state,
the $a \,^1\Delta_g$, and the $a \,^1\Sigma_g^+$ singlet excited states
using the MSM and various DFAs.  As already explained in Sec.~\ref{sec:theory},
our general procedure is to first construct
a reference state with half a spin $\alpha = \, \uparrow$ electron and half
a spin $\beta = \, \downarrow$ electron in each $\pi^*$ orbital and then
reoccupy the resultant orbitals to create the single-determinantal states 
(including that for the ground $X \,^3\Sigma_g^-$ state) needed
for the MSM calculation.  Otherwise the procedure is very similar to that
described in the previous subsection where symmetry breaking was discussed.
Our major focus will be on the behavior of different {\em functionals} (DFAs)
in MSM calculations of the first three electronic states of O$_2$.
The ground $X \,^3\Sigma_g^-$ state will be treated more summarily as
the quality of ground-state properties have been extensively studied as
a function of choice of DFA.  Our main interest is in the choice of DFA
for describing the two $^1$O$_2$ excited states using the MSM method as
very little seems to be known about this.

% ----------------------------------------------------------
\begin{table}
\caption{
\label{tab:jacob}
John Perdew's vision of a Jacob's ladder for DFAs \cite{PS01,PRC+09}. On the left of the ladder are the
new variables which may be used to make functionals once that rung is reached. On the right is the
name usually given to that level of approximation.  In the middle is a user drawn as an angel in 
	the smiley (emoji) approximation.  The user needs to be able to ascend and descend the ladder in search of the
right compromise between accuracy and computational resources to meet his or her needs.
}
\begin{center}
\begin{tabular}{c|c|c}
\multicolumn{3}{l}{Quantum Chemistry Heaven} \\
                     &        &              \\
$\psi_{\mbox{virt}}$ & \rule{2cm}{0.4pt} & double hybrid \\
                     &        &               \\
$\psi_{\mbox{occ}}$   & \rule{2cm}{0.4pt} & hybrid, mGGA hybrid \\
                     &        &               \\
		     & \LARGE $ \ddot{\smile}   $ \normalsize &        \\
$\tau$, $\nabla^2 \rho$ & \rule{2cm}{0.4pt} & mGGA \\
                     &        &              \\
$\nabla \rho$        & \rule{2cm}{0.4pt} & GGA \\
                     &        &              \\
$\rho$               & \rule{2cm}{0.4pt} & LDA \\
                     &        &              \\
\multicolumn{3}{l}{Hartree Theory Earth} \\
\end{tabular}
\end{center}
\end{table}
% ----------------------------------------------------------
Although repetitive for DFT experts, it seems appropriate to begin with a 
very brief reminder of the 
various rungs of the Jacob's ladder of DFAs shown in 
{\bf Table~\ref{tab:jacob}}.
The basic idea is that we climb the ladder by allowing our exchange-correlation energy
functional to depend on more and more variables.  This typically requires us to calculate
more complicated quantities, making more resource-intensive calculations, but is not
guaranteed to give better results.  However there is a tendency for functionals from higher
rungs of the ladder to be more accurate than those from lower rungs of the
ladder (see, e.g., Ref.~\cite{GM19}).   The lowest rung is the local (spin)
density approximation (LDA) where the exchange-correlation (xc) energy density
at each point in a molecule only depends upon the density at that point
in the molecule.  The next level is constituted by the generalized gradient
approximations (GGAs) where the xc energy density depends also on the gradient
of the density at that point.  Meta GGAs (mGGAs) go a step further by including
either an orbital dependence through the kinetic energy density $\tau$ or
via the Laplacian of the density.  Hybrid functionals include some contribution
from the exact (also known as Hartree-Fock) exchange as well as GGA exchange
and correlation.  In {\sc deMon2k}, there are also hybrid functionals which are
constructed using mGGAs.

So much has been learned for the various levels of Jacob's ladder for 
ground-state 
calculations that must already be familiar to most users of DFT, that we do not 
want to belabor our comparison of DFAs.  However we do need to recall a few
things about what is known from ground state calculations before commenting
on MSM DFT excited-state calculations.

% -------------------------------------------------------------
\begin{figure}
\begin{center}
\includegraphics[width=0.9\textwidth]{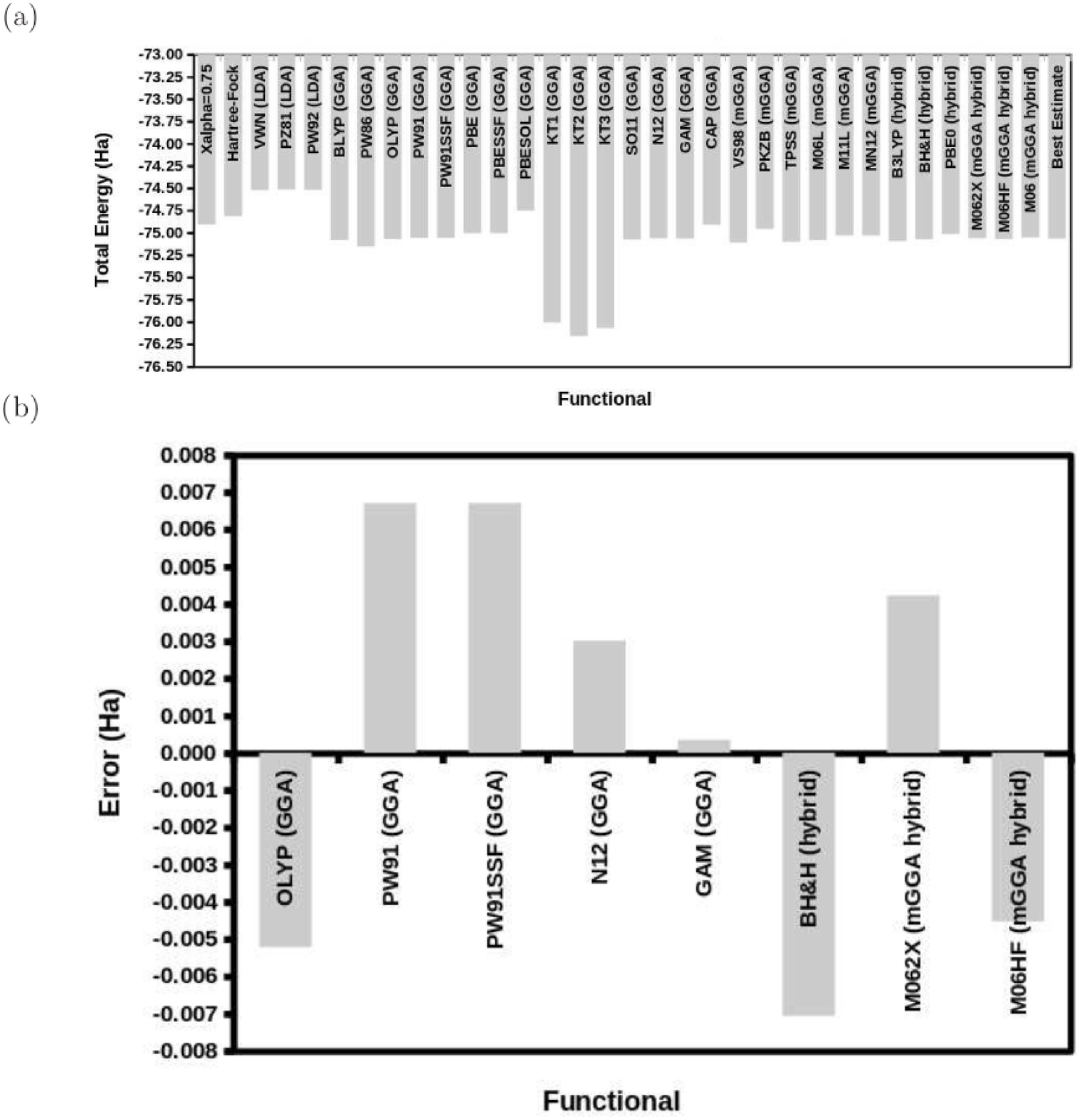}
\end{center}
\caption{\label{fig:atom}
Comparison of total O($^3P$) atomic energies calculated with 
various DFAs: (a) total energies, (b) functionals with an absolute error
of less than 0.2 eV with respect to the best estimate for the O($^3P$) 
atomic energy taken from Ref.~\cite{OG05}.
}
\end{figure}
% --------------------------------------------------------------
% {\bf Table~\ref{tab:atom}} and 
{\bf Figure~\ref{fig:atom}} compares calculations
of the total electronic energy of the O($^3P$) atom.  This quantity is needed
in order to evaluate the energy of the O$_2$ chemical bond in the three
electronic states.  The X$\alpha$ method was designed to behave like 
Hartree-Fock (HF) which gives too high an energy because of the lack of 
electron correlation.  Perhaps surprisingly the three LDA parameterizations 
over-estimate the atomic energies even worse than do X$\alpha$ and HF (leading to 
overbinding in molecules).  The GGAs, mGGAs, hybrids, and mGGA hybrids lead to 
substantial improvements over the LDA (X$\alpha$) and HF with, some notable 
exceptions such as PBESOL which was specifically parameterized to treat solids 
and the KT1, KT2, and KT3 functionals which were originally intended for 
nuclear magnetic resonance (NMR) calculations.  Only 5 GGAs (OLYP, PW91, PW91SSF, 
N12, and GAM), one hybrid (BH\&H), and two mGGA hybrids (M062X and M06HF) are 
within an arbitrary cut-off of 0.2 eV = 0.00735 Ha of our best estimate value.

% -------------------------------------------------------------
\begin{figure}
\begin{center}
\includegraphics[width=0.9\textwidth]{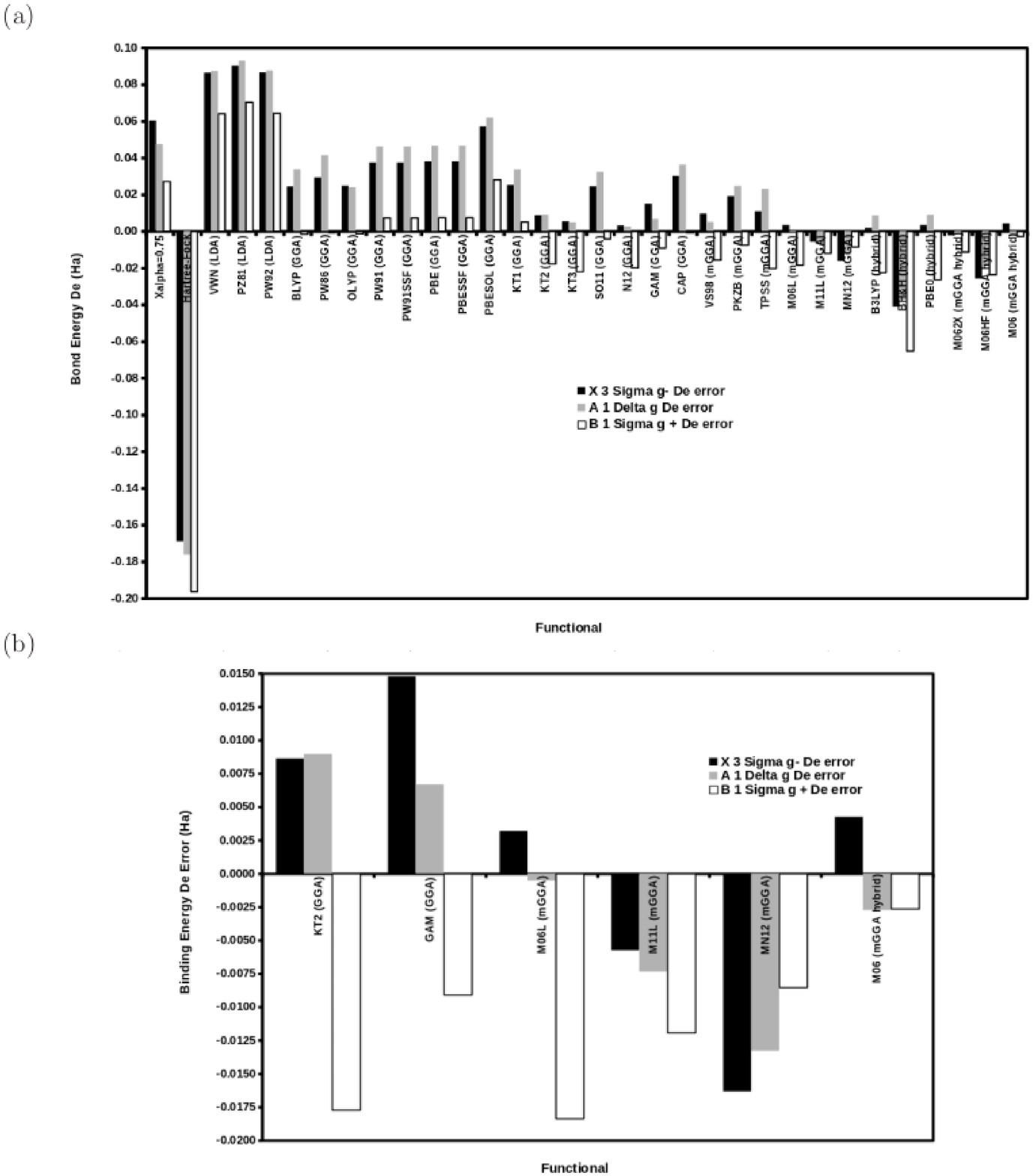}
\end{center}
\caption{\label{fig:BE}
Comparison of the binding energy, $D_e = 2 E(\mbox{O($^3P$)}) - E(\mbox{O$_2$})$,
calculated with various DFAs: 
	% (a) total binding energy (see the {\em Supplementary
 % Information} for a table of numerical values), 
	(a) error with respect
to the best estimate, and (b) zoom in on functionals showing the smallest errors.  
The best estimate is from Fig.~1 of Ref.~\cite{FCY+14}.
	(See the {\em Supplementary Information} for a table binding energies.)
}
\end{figure}
% --------------------------------------------------------------
It is often said that quantum chemistry works by error cancellation.  That is, that
errors in the energy of core electrons cancel out when chemical bond energies are
calculated.  This is why atomic O($^3P$) energies are not as important as the bond
energies shown in % {\bf Table~\ref{tab:BE}} and 
{\bf Fig.~\ref{fig:BE}}.  Before John Perdew spoke of Jacob's ladder, Axel Becke had 
already spoken of the three generations of DFAs.  
In Becke's scheme, X$\alpha$ and Hartree-Fock (HF) were pre-history.  We found severe
underbinding of O$_2$ at the HF level with respect to twice the energy of O($^3P$).
But this is the only functional where we see this problem.  X$\alpha$ was supposed to
be parameterized to behave like HF but actually behaves more like the LDA.
Becke's first generation is that of the LDA which was known to overbind molecules.
As we have seen in the case of O$_2$, this is because the LDA energy of O($^3P$) is
too high in energy.  The second generation is that of the GGAs which did much to correct
the overbinding problem and make DFT useful for chemical applications.  However GGAs 
could not yet approach chemical accuracy (defined as 1 kcal/mol = 0.001594 Ha).  Only 
the third generation --- namely hybrid functionals --- could approach this level of 
accuracy.  All of this is well-known for the ground state.  However we now see that the
same holds for the bond energies obtained for the $a \,^1\Delta_g$ excited state.  Oddly
enough most functionals give relatively small {\em absolute} errors for the $b \,^1\Sigma_g^+$
state.  However this is actually a bad thing as for most applications we are interested
in the energy of the $b \,^1\Sigma_g^+$ state relative to the $X \,^3\Sigma_g^-$ and
$a \,^1\Delta_g$ state.  Chosing an arbitrary cut-off of 0.02 Ha = 0.544 eV = 12.6 kcal/mol, 
we find that only two GGAs (KT2 and GAM), three mGGAs (M06L, M11L, and MN12), and one mGGA 
hybrid (M06) are able to obtain absolute energies to within this error range of our best 
estimate for the bond energies of all three states.

% -------------------------------------------------------------
\begin{figure}
\begin{center}
\includegraphics[width=0.9\textwidth]{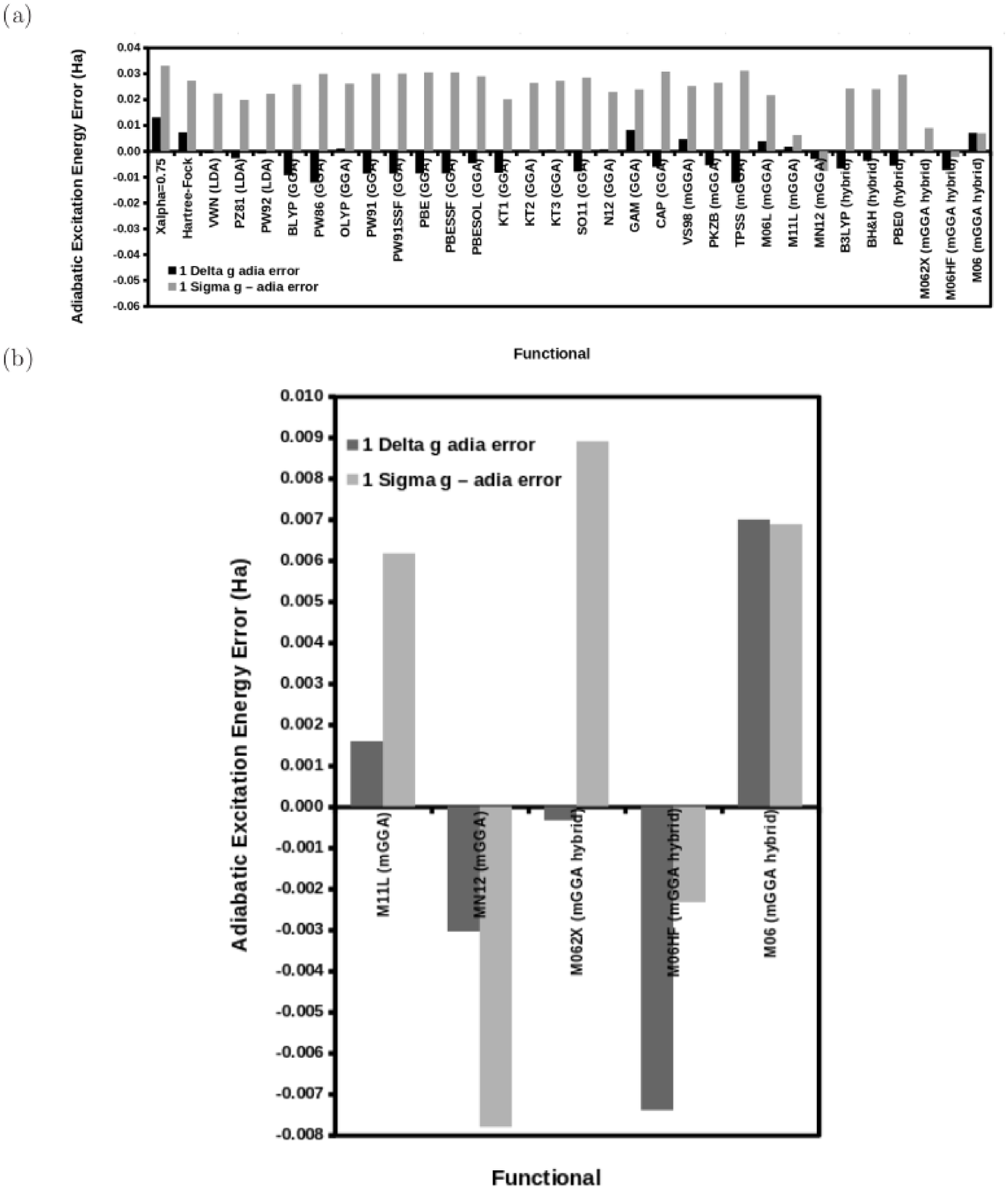}
\end{center}
\caption{\label{fig:exci}
Comparison of the singlet adiabatic excitation energies 
$\Delta E = D_e(\mbox{$^3$O$_2$}) - D_e(\mbox{$^1$O$_2$})$
calculated with various DFAs: % (a) total binding energy, 
(a) error with respect to the best estimate, and 
(b) zoom in on functionals showing the smallest errors.
The best estimate is from Fig.~1 of Ref.~\cite{FCY+14}.
}
\end{figure}
% --------------------------------------------------------------
A more delicate test of energetics is to calculate excitation energies.  Vertical
excitation energies are the difference of electronic energies of the excited states
and the ground state at the equilibrium geometry of the ground state, whereas 
adiabatic excitation energies are the difference of electronic energies of the
excited states and the ground state at the equilibrium geometry of the ground state.
As such, adiabatic excitation energies are the difference of the bond energies in the 
ground $^3$O$_2$ state and the two excited $^1$O$_2$ states.  Both types of excitation
energies are tabulated in the {\em Supplementary Information}.  Here, for a reason
which will be explained below, we will focus on adiabatic excitation energies 
({\bf Fig.~\ref{fig:exci}}).  It is evident that accurate $a \,^1\Delta_g$ excitation
energies are fairly easily calculated using the MSM method.  This means that it is
fairly easy to obtain a reasonably accurate spin-flip energy $F$.  In contrast,
most functionals significantly over-estimate the $b \,^1\Sigma_g^+$ excitation energy,
meaning that the spin-pairing energy $P$ is overestimated.  There are four exceptions
which stand out as being notably better in this respect, namely two mGGA (M11L
and MN12) and three mGGA hybrid (M062X, M06HF, and M06) functionals.
We have already noted that there is a scarcity of MSM DFT results for $^1$O$_2$
with which to compare but have already presented a comparison of the
results that we have found with some of our results in Table~\ref{tab:compare}.
These have already been discussed in the introduction.  A puzzling point is
why our M06HF/DEF2-TZVPP MSM results (see the excitation energies in the
{\em Supplementary Material}) are so different from the MSDFT/M06-HF/cc-pVTZ 
results reported in Ref.~\cite{Q20}.  We think that this might be due to
confusion between the M062X, M06HF, and M06 mGGA hybrid DFAs which were
all presented in the same reference \cite{ZT07} as our M06/DEF2-TZVPP MSM 
results certainly do resemble the MSDFT/M06-HF/cc-pVTZ results reported in 
Ref.~\cite{Q20}.

% -------------------------------------------------------------
\begin{figure}
\begin{center}
\includegraphics[width=0.9\textwidth]{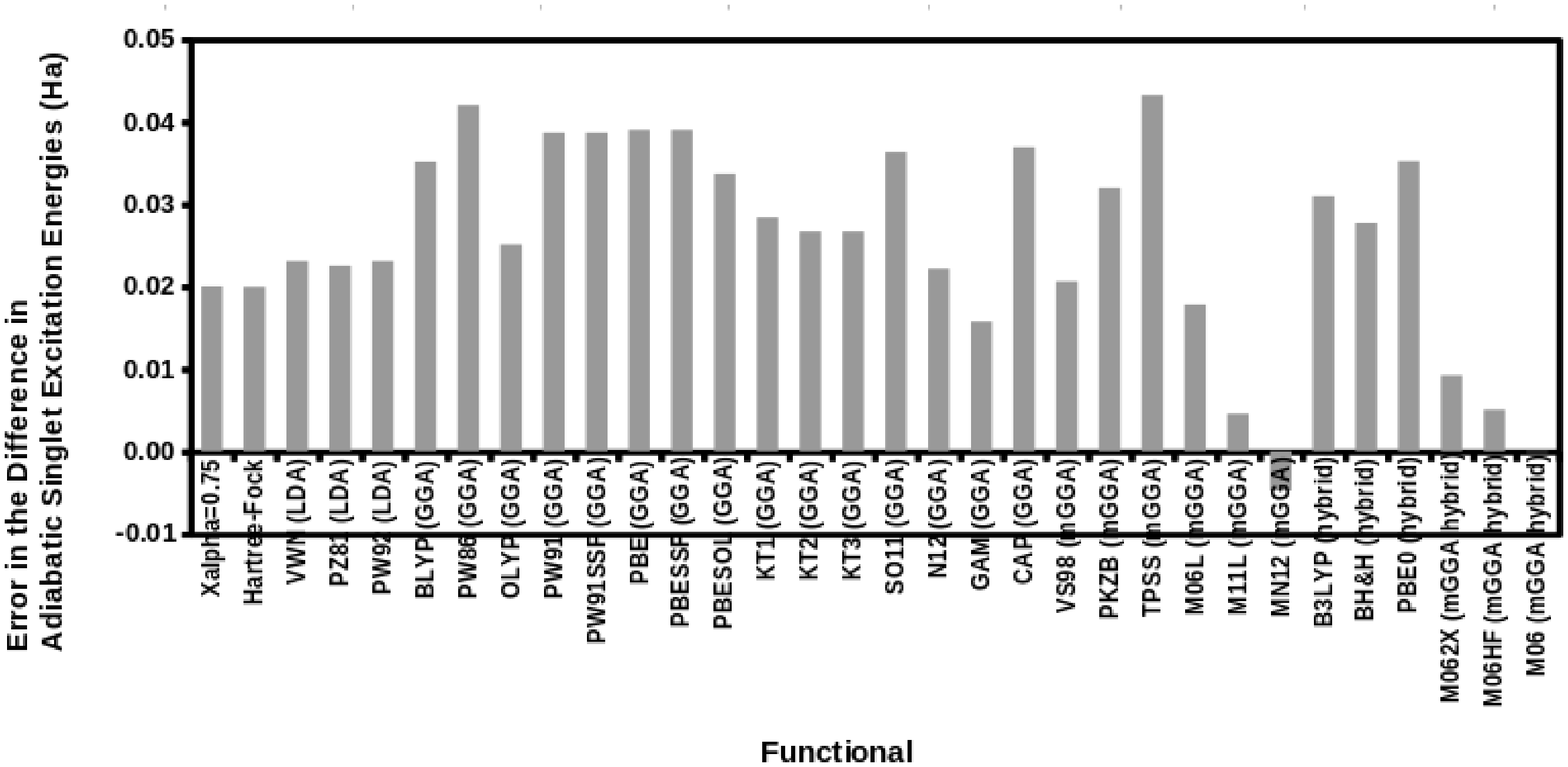}
\end{center}
\caption{\label{fig:PE}
Comparison of the error in twice the pairing energy 
$2\tilde{P} = E[b \,^1\Sigma_g^-] - E[a \,^1\Delta_g]$
calculated with various DFAs.
}
\end{figure}
% --------------------------------------------------------------
Some time ago, one of us (MEC) was involved in an extensive series of studies comparing
DFAs for calculating the $^5T_{2g}: (t_{2g})^4 (e_g)^2$-$^1A_{1g}:(t_{2g})^6 (e_g)^0$ 
spin-crossover energy difference in octahedral Fe(II) complexes 
\cite{FMC+04,FCL+05,GBF+05,LVH+05,ZBF+07}.  An important conclusion of these studies
is that HF would tend to stabilize the configuration $\pi_x [\uparrow \downarrow] [\,\,\,\, \,\,\,\,] \pi_y$ more than would the LDA or GGAs with respect to the configuration
$\pi_x [\uparrow \,\,\,\,] [\uparrow \,\,\,\,]$.  That is, HF should underestimate the
spin-pairing energy $\tilde{P} = E(\pi_x [\uparrow \downarrow] [\,\,\,\, \,\,\,\,] \pi_y) - 
E(\pi_x [\uparrow \,\,\,\,] [\uparrow \,\,\,\,])$ while the LDA and
GGAs tend to overestimate $\tilde{P}$.  We also found that the OLYP functional tended to
give a less severe overestimation.  This is confirmed in {\bf Fig.~\ref{fig:PE}}.  Note
that $\tilde{P} \neq P$ represent two different pairing energies.  

Returning again to the problem of octahedral Fe(II) spin-crossover complexes, it has 
been argued \cite{SKS+18} that the real problem lies not with the total exchange-correlation 
energy but with the accuracy of the exchange-correlation potential which is unable to produce 
an accurate enough density.  They show that the problem disappears when an accurate density 
is used in the xc energy functional rather than a self-consistent density.  Although we find this
to be an interesting discovery, we have not pursued it for the present problem.

% -----------------------------------------------------
\subsection{Shape of the Multiplet Sum Method Surfaces}
% ------------------------------------------------------

Thus far we have focused mainly on energy differences (namely bond energies and
excitation energies).  We now wish to go further and discuss the
shape of the PECs.  As the $^3$O$_2$ ground state and the two $^1$O$_2$ excited states differ
only by a redistribution of the two $\pi^*$ electrons, it might be anticipated that the
$X \,^3\Sigma_g^-$, $a \,^1\Delta_g$, and $b \,^1\Sigma_g^+$ PECs should all have roughly
the same shape.  However further reflection shows that this cannot occur if they have 
different bond energies ($D_e$) and the same dissociation limit.  There must be some
state-dependent shape differences and the question arises as to what extent this is reflected
in our DFT MSM calculations.  All our MSM curves are derived from the same reference state
with half a spin $\alpha = \, \uparrow$ and half a spin $\beta = \, \downarrow$ electron in
each $\pi^*$ orbital.  Once upon a time, such a state would have been indistinguishable from
the ground state because exchange-correlation energy functionals had no spin dependence.
However that was a long time ago.  Nevertheless we might expect that the different PECs
should behave something like the ground $X \,^3\Sigma_g^-$ PEC, particularly since no
additional self-consistent field (SCF) cycles are carried out after reoccupying the orbitals
of the reference state.  This is indeed what is seen in the graphs shown in the {\em Supplementary
Information}.  But let us take a closer look.

% -------------------------------------------------------------
\begin{figure}
\begin{center}
\includegraphics[width=0.9\textwidth]{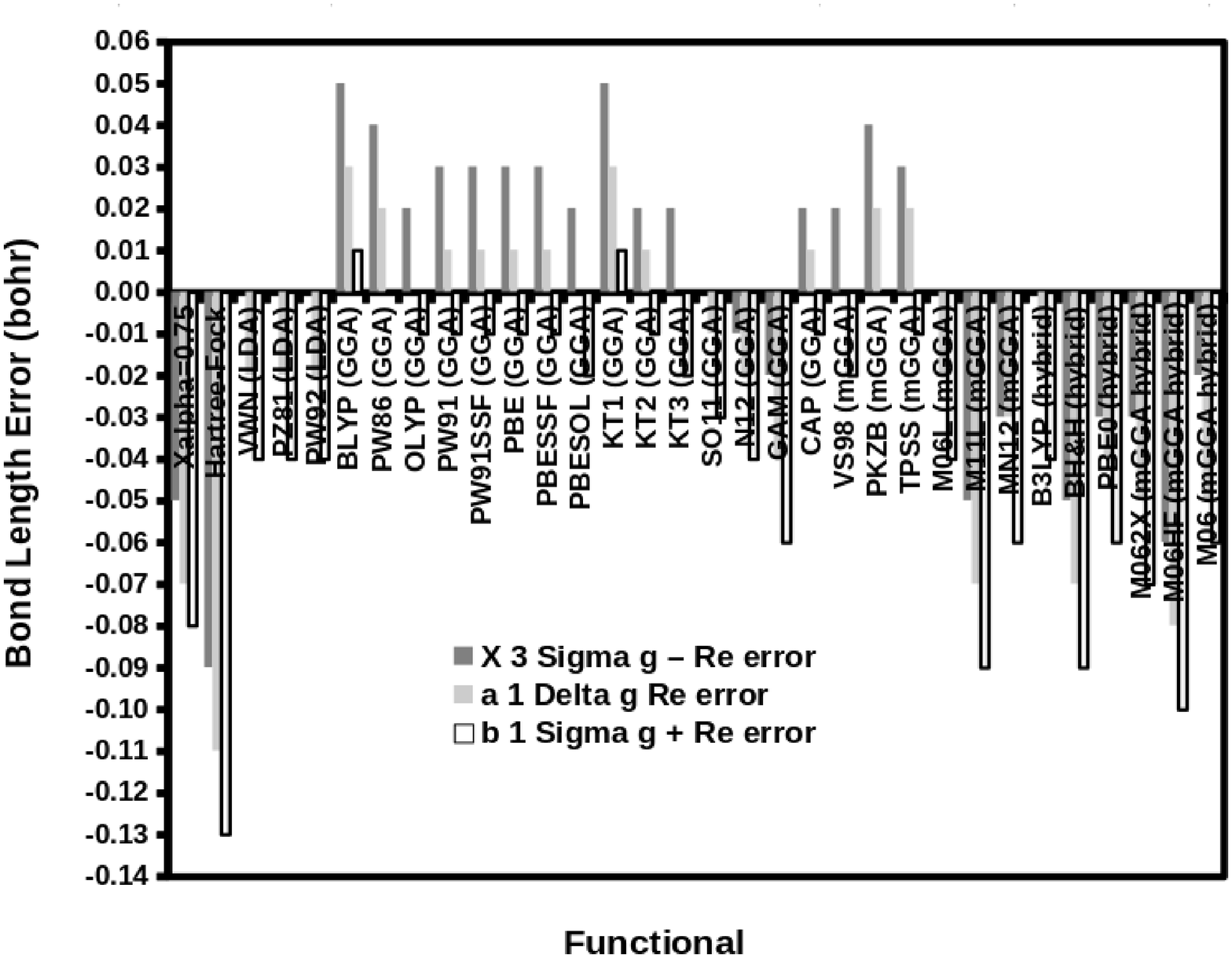}
\end{center}
\caption{\label{fig:dist}
Bond distances calculated with various DFAs compared with best estimate
values from Refs.~\cite{FCY+14,NISTASD}: % (a) distance and (b) 
	error
in the distance with respect to Ref.~\cite{NISTASD}.
}
\end{figure}
% --------------------------------------------------------------
{\bf Figure~\ref{fig:dist}} confirms that the MSM DFT
bond lengths with a given functional seem to be roughly independent of the
electronic state.  This is in contast with the experimental observation \cite{FCY+14,NISTASD}
that the higher-energy electronic states which have smaller bond energies also
have longer bond lengths.  Moreover, the functionals which were previously found
to be optimal for energies are not the ones found to give the best bond lengths.

Let us turn next to the curvature at the bottom of the PEC wells as measured by the harmonic
frequency.  As may be expected from the above discussion of symmetry-broken and 
symmetry-unbroken problems and from the graphs in the {\em Supplementary Information}, 
our symmetry-unbroken PECs are expected to be significantly more parabolic, 
\begin{equation}
  V(R) = \frac{k}{2} \left(R-R_e \right)^2 + \mbox{HOT} \, ,
  \label{eq:results.2}
\end{equation}
than the best estimate PECs.  Here ``HOT'' stands for ``higher-order terms.''  
To find $k$ and $a$, we fit three points around the minimum to the quadratic,
\begin{equation}
  V(R) = A R^2 + B R + C \, .
  \label{eq:results.3}
\end{equation}
Then $A = k/2$.  Also from the classical mechanics of a simple harmonic oscillator,
$k=\mu \omega_0$, where the reduced mass $\mu = m^2/(m+m) = m/2$ in terms of the actual
mass of an oxygen atom
$m = M/N_A = (\mbox{15.999 g/mol})/(\mbox{6.022 $\times$ 10$^{23}$/mol}) = 2.6568 \times 10^{-23} 
\mbox{ g} = 29165 \, m_e $.  So 
\begin{equation}
  \hbar \omega_0 = 2 \sqrt{\frac{A}{m}} \, .
  \label{eq:results.4}
\end{equation}
% -------------------------------------------------------------
\begin{figure}
\begin{center}
\includegraphics[width=0.9\textwidth]{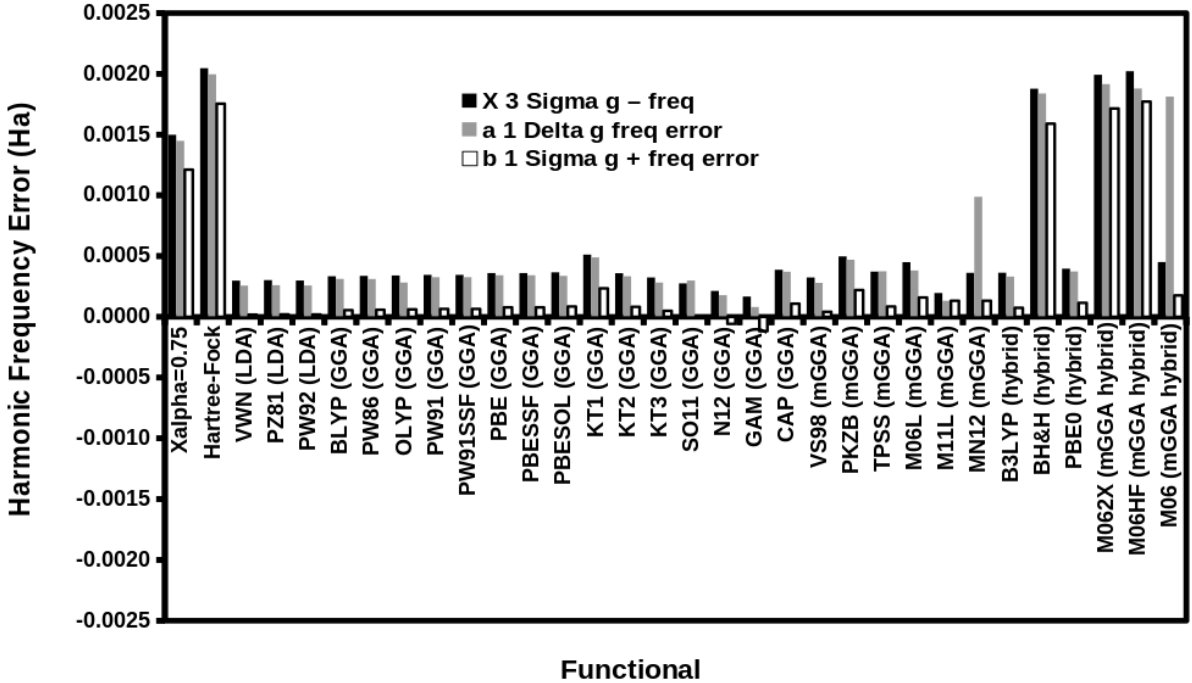}
\end{center}
\caption{\label{fig:freq}
Harmonic frequencies calculated with various DFAs compared with best estimate
values from Refs.~\cite{FCY+14,NISTASD}: % (a) frequency and (b) 
	error in the
frequency with respect to values from Ref.~\cite{FCY+14}.
}
\end{figure}
% --------------------------------------------------------------
Harmonic frequencies are shown in {\bf Fig.~\ref{fig:freq}} (and numbers are tabulated
in the {\em Supplementary Information}).  We might expect the harmonic frequencies to
decrease as we go to higher-energy states because the PEC well is becoming shallower
and this is indeed what is seen in the experimental data of Ref.~\cite{NISTASD}.
However the effect is small and we do not see it in the harmonic frequencies extracted
from Fig.~1 of Ref.~\cite{FCY+14}.  The LDA is famous for giving good harmonic frequencies
(due to error cancellation) and this is clearly seen in Fig.~\ref{fig:freq}.  That we
also see this for the LDA, GGAs, and mGGAs, both for the ground and for the two excited
states is very reassuring.  Less reassuring are the relatively large errors seen for the
MN12 mGGA and for the M06 mGGA hybrid for the $a \,^1\Delta_g$ state and for all the states
for the BH\&H hybrid, the M062X mGGA hybrid, and the M06HF hybrid.

We may also estimate anharmonicities by assuming that the PECs may be approximated, at
least roughly, by a Morse potential \cite{M29},
\begin{eqnarray}
  V(R) & = & D_e e^{-(R-R_e)} \left( e^{-a(R-R_e)} - 2 \right) \nonumber \\
  & = & D_e \left[ a^2 \left( R-R_e \right)^2 - 1 \right] + \mbox{HOT} \,
  \label{eq:results.5}
\end{eqnarray}
As the Schr\"odinger equation may be solved exactly for a Morse potential to obtain,
\begin{eqnarray}
	E_n & = & \hbar \omega_0 \left( n + \frac{1}{2} \right) - 
     \frac{\left[ \hbar \omega_0 \left( n+ \frac{1}{2} \right) \right]^2 }{4 D_e} \nonumber \\
	& = & \hbar \omega_0 \left( n + \frac{1}{2} \right) + \hbar x_0 \omega_0 \left( n+ \frac{1}{2} \right)^2 \, ,
  \label{eq:results.6}
\end{eqnarray}
we may also obtain the anharmonicity from the bond energy and harmonic frequency as,
\begin{equation}
  \hbar x_0 \omega_0 = \frac{\hbar^2 \omega_0^2}{4 D_e} \, .
  \label{eq:results.7}
\end{equation}
For completeness, we note that,
\begin{equation}
  a = \frac{\omega_0}{2} \sqrt{\frac{m}{D_e}} \, ,
  \label{eq:results.8}
\end{equation}
in the Morse equation so that we may plot the PECs.
% -------------------------------------------------------------
\begin{figure}
\begin{center}
\includegraphics[width=0.9\textwidth]{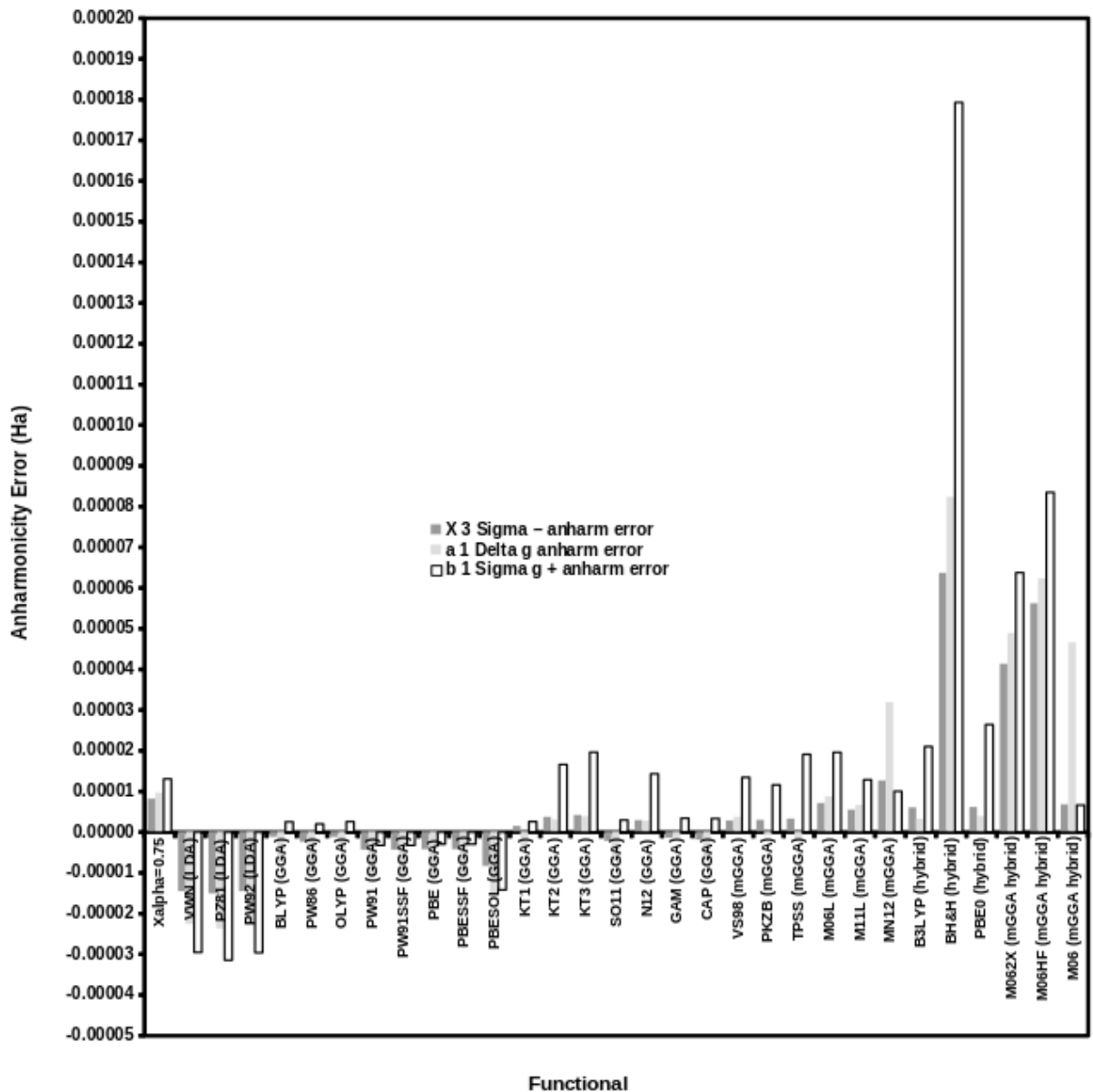}
\end{center}
\caption{\label{fig:anharm}
Anharmonic frequency corrections calculated with various DFAs compared with 
best estimate values from Refs.~\cite{FCY+14,NISTASD}: 
	% (a) anharmonic correction and (b) 
	error in the anharmonic correction with respect to values from 
Ref.~\cite{FCY+14} (except for HF which shows much larger errors).
}
\end{figure}
% --------------------------------------------------------------
According to Eq.~(\ref{eq:results.7}), the anharmonicity is determined by the
harmonic frequency and by the bond energy.  {\bf Figure~\ref{fig:anharm}} shows
that the best functionals for calculating the anharmonicity are also pretty much the best
ones for calculating the harmonic frequency.

% -------------------------------------------------------------
\begin{figure}
\begin{center}
\includegraphics[width=0.9\textwidth]{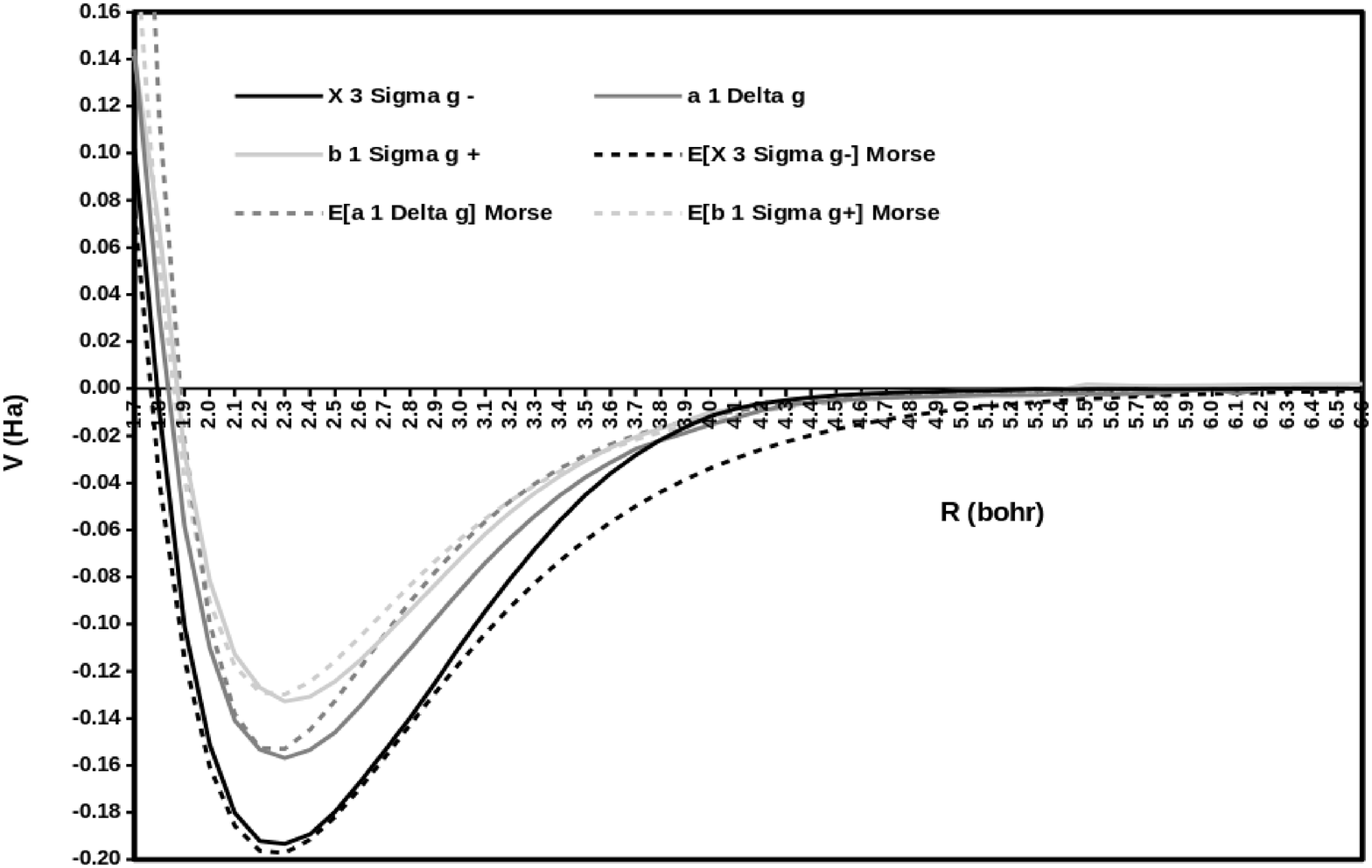}
\end{center}
\caption{\label{fig:M06Morse}
Comparison of the Morse potential calculated using data obtained via MSM
calculations with the M06 GGA and best estimate PECs from Ref.~\cite{FCY+14}.
}
\end{figure}
% --------------------------------------------------------------
The question now arises as to how to choose the best functional for DFT MSM
calculations.  There are, in fact, two questions: (a) ``What functional is best
for a random unknown molecule?'' and (b) ``What functional is best for O$_2$?''
Clearly we can only answer question (b) and even in that case, it depends upon
making choices and compromises between the shapes of the PECs and their absolute
or relative energies.  {\bf Figure~\ref{fig:M06Morse}} shows the Morse curves
obtained from the MSM with the M06 GGA which seems to be a fair compromise
between competing interesting features of the PECs, albeit one which places more
emphasis on absolute energies than upon other factors.

%%%%%%%
% EOF %
%%%%%%%
% -----------------------------------------------
\section{Conclusions and Perspectives}
\label{sec:conclude}

Twenty percent of our atmosphere is made up of molecular oxygen, which is
nearly all in the ground $X \,^3\Sigma_g^-$ state.  Some of this
relatively kinetically-unreactive triplet molecule ($^3$O$_2$) is constantly being
converted into the very reactive singlet form ($^1$O$_2$).  Understanding how this
happens, how to take advantage of $^1$O$_2$ to carry-out desired reactions, and how
to get rid of $^1$O$_2$ to avoid undesirable reactions involves understanding how
$^1$O$_2$ reacts with large molecules, such as C$_{60}$.  The size of these molecules
tends to rule out the use of sophisticated, accurate, yet computer-resource intensive, 
{\em ab initio} methods in favor of the computationally simpler density-functional
theory (DFT) approach.  But it is this very simplicity which should act as a warning
when applying DFT to intrinsically-multideterminantal problems such as open-shell
systems, excited states, and the calculation of potential energy curves (PECs) that
involve the making and breaking of bonds.

Hohenberg-Kohn-Sham DFT \cite{HK64,KS65} assumes the existence of a fictitious system
of noninteracting electrons whose ground-state charge density is the same as that of
the real system of interacting electrons.  This is the assumption of noninteracting 
$v$-representability which does not always hold \cite{CH12}, particularly
in strongly multideterminantal situations such as open-shell molecules or 
when bonds are being made or broken.  Requiring effective noninteracting
$v$-representability of an approximate density functional means placing
particularly difficult demands on density-functional approximations (DFAs)
as it means that we expect them to be able to model the behavior of 
an intrinsically multideterminantal wave function, which even the exact
functional might not be able to model, using only a single-determinantal
Kohn-Sham wave function.  This is why the Ziegler-Rauk-Baerends-Daul 
\cite{ZRB77,D94} multiplet sum method (MSM)
proposes an alternative way of treating intrinsically multiderminantal systems by using
symmetry to deduce equivalent energy expressions involving only single-determinantal energy
expressions.  This method is widely used to treat open-shell singlet excited states.

We have shown that the $a \,^1\Delta_g$ state of $^1$O$_2$ may be treated using exactly the
same DFT MSM approach used for open-shell singlet excited states.  Oddly enough this does
not seem to be widely known (or at least we have been able to find at most one other reference
using something like this approach for $^1$O$_2$ \cite{Q20}).  We speculate that this 
might be due to the large amount
of confusion in the photochemical literature regarding the proper molecular orbital picture
of the $^1$O$_2$ excited states combined, perhaps, with little or no overlap between the
transition-metal complex and magnetic property communities most likely to use the DFT MSM
approach and the community interested in $^1$O$_2$ photosensitizers.  Furthermore, we point
out that the DFT MSM may also be used to obtain the $b \,^1\Sigma_g^+$ state of $^1$O$_2$.

The main limitation is that the symmetry analysis underlying the MSM and the labeling of 
states in small molecules is incompatible with the symmetry-breaking 
different-orbitals-for-different-spins (DODS) method typically used to study PECs over
a broad range of bond distances.  Instead we have made an extensive study of the behavior
of the DFT MSM for describing PECs in the region close to the equilibrium bond length
where no symmetry breaking occurs.  We found good $a \,^1\Delta_g$ excitation energies
with most functionals but that obtaining accurate ground- and excited-state bond energies
and a good $b \,^1\Sigma_g^+$ excitation energy requires us to climb the Jacob's ladder of 
functionals to rungs using mGGAs and mGGA hybrid functionals. Unfortunately this is also
accompanied by some degradation of the quality of bond lengths and harmonic frequencies.
Nevertheless it is possible to obtain reasonable approximate PECs for all three states 
($X \,^3\Sigma_g^-$, $a \,^1\Delta_g$, and $b \,^1\Sigma_g^+$) by the use of a Morse
potential extrapolation to longer bond lengths.  Although a somewhat arbitrary choice,
we illustrated how this works with the M06 functional.

This brings us to the point of how to move beyond the MSM to be able to apply
DFT-based methods to increasingly complicated photochemical systems.  Some
readers may recognize that one of the authors (MEC) was a pioneer in the
formulation of time-dependent (TD) DFT for treating excited states in
quantum chemistry \cite{C95} (see also the reviews in Refs.~\cite{C09,CH12}).
An obvious problem with the MSM is that it relies too much on symmetry
to estimate the trial wave function which makes it difficult to apply
to larger, less symmetric systems.  The MSM also requires human intervention
to chose the reference system and the configurations to be treated.  In
this sense, the MSM resembles multiconfigurational wave function methods
such as the complete active space (CAS) self-consistent field (SCF) method.
In contrast TD-DFT provides a fully automatic procedure for treating 
excited states which meshes nicely with algorithms already needed in DFT 
programs to evaluate gradients and response properties.  This, and its
good performance to computer resource ratio, is one of the reasons that
TD-DFT has become a major tool in the quantum chemistry toolbox.

But it would be very wrong to think that TD-DFT replaces the MSM.  Indeed
MEC has always found it remarkable that, though the MSM and TD-DFT are
based upon very different formalisms, they typically give very similar
answers {\em in situations where both apply}, which typically means 
(for TD-DFT) situations where the ground state is reasonably well described 
by a single determinantal wave function and excitations are limited to
energetically low-lying localized states with little or no relaxation of
the charge density and (for the MSM) there is a good reference state
and enough symmetry relations may be deduced to reduce state energies to
weighted sums of single determinantal energies.  It is particularly 
fascinating that the TD Hartree-Fock (HF) Tamm-Dancoff approximation (TDA) 
and $\Delta$SCF (HF) energy expressions, neglecting relaxation, are
identical, but that this same statement is not true for TD-DFT and the $\Delta$SCF DFT
method.  Early work to reconcile these two methods was reported in
Refs.~\cite{C99b} and \cite{CGG+00}.  It has been continued in some lovely
creative work on constricted variational DFT (CV-DFT) \cite{ZSK+09,ZKC12} 
that originated in the group of Ziegler (now deceased).

Given that the MSM and TD-DFT approaches are not as independent as they
might seem, that both methods (and combinations thereof) are continuously
evolving, and that a conclusion is a good place to give a few perspectives
for the future but is {\em not} the proper place for a review, then we will try to
limit our perspectives primarily to how the MSM method has been improved
for applications to more complex problems.  Nevertheless we cannot resist
pointing out that, beginning with early work on TD-DFT (often with {\sc deMon2k}) 
for open-shell systems \cite{HH99,GCS00,CIC06,ICJC09,MC17}, spin-flip TD-DFT 
\cite{SK03,SHK03,WZ04,WZ06,MG09,HNI+10}, and dressed TD-DFT 
\cite{CZMB04,MZCB04,C05,HIRC11,CH15}, progress is also being 
made towards a more versitile TD-DFT able to handle more complex
photochemical problems.  [The list of cited references is grossly incomplete,
but (as said above) this is no place for a review!]  Particularly relevant
to the present paper is the work of Guan {\em et al.} applying TD spin-flip DFT 
to the excited-state PECs of O$_2$ \cite{GWZC06}. 

Let us now return to and focus specifically on evolutions of the MSM.
The natural evolution is to expand the active space and use MSM {\em and other}
techniques to evaluate the matrix elements of a small configuration interaction (CI)
problem.  Historically MSM practitioners would use symmetry to write down as many
relations as possible and attempt to express all needed CI matrix elements in terms
of single-determinantal energies.  When this was not possible, then a few additional
electron repulsion integrals (ERIs) would be evaluated explicitly using the DFT orbitals.
In recent years, some new tricks have been added and the resultant improved MSM goes
by names like multistate DFT (MSDFT) and constrained DFT configuration interaction 
(CDFT-CI).  Let us try to give the gist of these two methods.  Both methods seem to
be inspired to some extent by Marcus' theory of charge transfer and, in particular,
resemble the creation of configuration mixing of delocalized (adiabatic) states from 
noninteracting localized (diabatic) states.  The relation with the MSM enters almost
accidently and with little or no discussion.

In MSDFT \cite{CSMG09,RPB+16,AAG+17,GCLG17,GQT+17,YGRG19,Q20}, also called valence-bond
DFT (VBDFT) \cite{CSMG09}, the diabatic states are based upon reactant and product orbitals
and coupling is based upon arguments similar to Marcus-Hush theory \cite{CSMG09}.
The MSM was introduced in Ref.~\cite{GCLG17} in order to be able to handle spin multiplets
and was also used in subsequent papers (such as Ref.~\cite{Q20}).

CDFT-CI comes from the group of Van~Voorhis \cite{WV06,WCV07,KV10,KKV12,MV15}.  
It makes no reference to the MSM, but nevertheless bears a strong resemblance to MSDFT.
CDFT-CI applies a constraint to localize spin and changes on user-defined
parts of the system which may not otherwise be very well described by DFT.  Diagonal
elements of the CI matrix are (single-determinantal) DFT energies from the constrained
states and off-diagonal elements are obtained by an essentially exact expression obtained
using the constraining potential.  This method is implemented in {\sc deMon2k} 
\cite{DS10,RLDD12,DAH+19}.  

So, in our O$_2$ case, the two $\pi^*$ orbitals could be considered as subsystems to 
which the constraints should be applied (though this to our knowledge has never yet 
been tried).  
It should then be possible to go further and use either MSDFT or CDFT-CI (or both) to consider
what happens when $^1$O$_2$ reacts with a molecule (such as C$_{60}$).  Of course the 
presence of the molecule will lead to the loss of degeneracy of the $a \,^1\Delta_g$ state
and we could then begin to test the empirical rules given by Kearns (p.~414 of Ref.~\cite{K71})
for $^1$O$_2$ reactions and the qualitative ways of thinking about diradical reactions 
advocated by Hoffmann and coworkers \cite{SCZ+19}.  In particular, when is it better to think
in terms of localized closed-shell singlets and when is it better to think in terms of
delocalized open-shell singlets (i.e., a little like in Figs.~\ref{fig:MOpi_xy} and 
\ref{fig:MOpi_pm})?

%%%%%%%
% EOF %
%%%%%%%
% ---------------------------------------
\section*{Acknowledgements}
\label{sec:thanks}

AJE, OM, and MEC met through the African School of Electronic Structure 
Methods and Applications (ASESMA) and are grateful for the learning and
networking possibilities provided by ASESMA.  MEC been involved in the
deMon developers group from the time of its creation.  
MEC is grateful to Pierre Girard for configuring the personal computer on
which some of the calculations reported here were performed.

%%%%%%%
% EOF %
%%%%%%%
% --------------------------------------
\section*{Supplementary Information}
\label{sec:SI}

The following supplementary information associated with this
article is available on-line:
\begin{enumerate}
  \item Digitized reference potential energy curves
  \item Group theoretic analysis
  \item Sample {\sc deMon2k} input
  \item Multiplet sum method potential energy curves
  \item Tables comparing calculated and experimental atomic and 
	diatomic calculated parameters
  \item Author contributions
\end{enumerate}

%%%%%%
% EOF
%%%%%%
% ============================================================
% \newpage
% \UN
% \bibliographystyle{myaip}
% \bibliography{refs}

% % ----------------------------------------------------------
\end{document}